\documentclass[fleqn,usenatbib]{mnras}

\usepackage{newtxtext,newtxmath}

\usepackage[T1]{fontenc}
\usepackage{ae,aecompl}

\usepackage{graphicx}	
\usepackage{amsmath}	
\usepackage{amssymb}	

\usepackage{pdflscape}
\usepackage{hyperref}
\usepackage{microtype}
\usepackage{color}
\usepackage[utf8]{inputenc}

\newcommand{\HI}{H$\,${\sc i}}

\newcommand{\mathHI}{{\mbox{\scriptsize \HI}}}

\graphicspath{{./}{figures/}}

\title[Low-$z$ Power Spectrum]{\huge
The Power Spectrum of the Lyman-$\alpha$ Forest at $z < 0.5$
}
\author[Khaire et al.]
{
\parbox{\textwidth}{
Vikram Khaire,$^{1}$\thanks{E-mail:vkhaire@physics.ucsb.edu} Michael Walther,$^{1,2,3}$ Joseph F. Hennawi,$^{1,2}$ Jose O\~norbe,$^{4}$ Zarija Luki\'c,$^{5}$ 
J. Xavier Prochaska,$^{6}$ Todd M. Tripp,$^{7}$ Joseph N. Burchett$^{8}$ and Christian Rodriguez$^{1,9}$
} 
\vspace*{10pt}\\
$^{1}$Physics Department, Broida Hall, University of California Santa Barbara, Santa Barbara, CA 93106-9530, USA\\
$^{2}$Max-Planck-Institut für Astronomie, Königstuhl 17, 69117 Heidelberg, Germany\\
$^{3}$International Max Planck Research School for Astronomy \& Cosmic Physics at the University of Heidelberg, Germany\\ 
$^{4}$Royal Observatories, Blackford Hill, Edinburgh EH9 3HJ, UK\\ 
$^{5}$Lawrence Berkeley National Laboratory, 1 Cyclotron Road, Berkeley, CA 94720, USA\\
$^{6}$Astronomy and Astrophysics, UC Santa Cruz, 1156 High St., Santa Cruz, CA 95064, USA\\ 
$^{7}$Department of Astronomy, University of Massachusetts - Amherst, 710 North Pleasant Street, Amherst, MA 01003-9305, USA\\
$^{8}$UCO / Lick Observatory, University of California, Santa Cruz, CA 95064, USA\\ 
$^{9}$Department of Physics and Astronomy, University of California, Irvine, CA 92697, USA
}   
\date{Accepted 2019 January 29. Received 2019 January 14; in original form 2018 September 4}
\pubyear{2018}
\begin{document}
\label{firstpage}
\pagerange{\pageref{firstpage}--\pageref{lastpage}}
\maketitle





\begin{abstract}
We present new measurements of the flux power-spectrum $P(k)$ of the $z<0.5$ \HI\ Lyman-$\alpha$ 
forest spanning scales $k \sim 0.001-0.1\, \mathrm{s \, km}^{-1}$. These results were derived from 
65 far ultraviolet quasar spectra (resolution $R \sim 18000$) observed with the Cosmic Origin 
Spectrograph (COS) on board the Hubble Space Telescope.
The analysis required careful masking of all contaminating,
coincident absorption from \HI\ and metal-line transitions of
the Galactic interstellar medium and intervening absorbers as well as proper treatment of the 
complex COS line-spread function.
From the $P(k)$ measurements, we estimate the \HI\ photoionization rate
($\Gamma_{\rm HI}$)  in the $z<0.5$ intergalactic medium.
Our results confirm most of the previous $\Gamma_{\rm HI}$
estimates.
We conclude that previous concerns of a photon underproduction
crisis are now resolved by demonstrating that the measured $\Gamma_{\rm HI}$ 
can be accounted for by
ultraviolet emission from quasars alone.
In a companion paper, we will present constraints on the thermal state of  the $z<0.5$ 
intergalactic medium from the $P(k)$ measurements presented here.

\end{abstract}


\section{Introduction} \label{sec:intro}
The intergalactic medium (IGM), being the largest reservoir of the baryons in the Universe, plays 
an important role in the formation of cosmic structures. 
The ultraviolet (UV) radiation emanating from this cosmic structure photoionizes and heats the IGM. 
The trace amount of neutral hydrogen in the highly-ionized IGM imprints a swath of absorption lines
on the spectra of background quasars known as the Lyman-$\alpha$ forest. 
Observations of the Lyman-$\alpha$ forest in a large sample of background quasar sightlines can 
probe the underlying density fluctuations in the IGM, measure its thermal state,  
and determine the amplitude of the UV ionizing background.

The temperature and density of the photoionized IGM follow a
tight power-law relation over two decades in the density,
$T(\Delta)=T_0\,\Delta^{\gamma -1}$, where $\Delta=\rho/\bar{\rho}$ is
the overdensity, $T_0$ is the temperature at mean density $\Delta=1$,
and $\gamma$ is the power-law index.  This power-law relation
quantifies the thermal state of the 
IGM \citep{Hui97,Theuns98,McQuinn16}. While a wide variety of statistics
have been applied to Lyman-$\alpha$ forest spectra with the goal of
measuring its thermal state \citep{Haehnelt98, Schaye99, Theuns00,
Zaldarriaga01, McDonald06,Lidz10, Becker11t, Bolton12, Rorai17,
  Hiss18}, the power spectrum of the transmitted flux is appealing for
several reasons: 
1) it is sensitive to a broad range of scales, in
particular, the small-scales that encode information about the IGM thermal
state, 2) it is thus 
capable of breaking strong parameter degeneracies, 3) systematics due to noise, metal-line
contamination, resolution effects, and continuum errors impact it in 
well-understood ways, and 4) it can be described by a simple multivariate
Gaussian likelihood enabling straightforward principled statistical
analysis and parameter inference \citep{Irsic17,Walther18a,Walther19}.
For these reasons, the power-spectrum has been used to constrain
parameters such as the UV ionizing background intensity \citep{Gaikwad17a}, alternate
cosmology models with warm and fuzzy dark
matter \citep{Viel08,Viel13,Garzilli17,Irsic17}, and cosmological parameters including
neutrino masses \citep{McDonald06,Palanque13ps,PD15Neutrino,Yeche17Neutrino,Irsic17t}.

There are many measurements of the flux power-spectrum at high redshifts \citep[e.g][]{McDonald00, 
Croft02, Kim04ps, Palanque13ps, Irsic17,Yeche17Neutrino, Walther18a} where ground-based telescopes 
with medium or high resolution spectrographs were used to observe the Lyman-$\alpha$ forest 
redshifted to optical wavelengths. 
However to date, there are no measurements at low-redshifts $z < 1.6$ 
\citep[but see][]{Gaikwad17a} where space-based observations are required because the redshifted
Lyman-$\alpha$ transition lies in the UV below the atmospheric cutoff. 
Recently,  large surveys \citep[e.g,][]{Tumlinson13, Danforth14, Burchett15, Borthakur15} have 
gathered a significant amount of Lyman-$\alpha$ forest spectra using the Cosmic Origin Spectrograph
(COS) on-board the Hubble Space Telescope (HST) that can be used to measure the Lyman-$\alpha$ 
forest power-spectrum at low redshifts. 

The power spectrum at low redshifts is of particular interest since it provides another method
for measuring the UV ionizing background, whereas previous work based on 
fitting the distribution of
column densities argued for a `photon underproduction crisis' \citep{Kollmeier14,Wakker15}. 
Also, it can measure
the thermal state of the low redshift IGM 
where long after the impulsive photoheating from
reionization events is complete, theory robustly predicts the IGM should have cooled
down to temperatures of $T_0\simeq 5000$ K at $z = 0$ \citep[see e.g][]{McQuinn16, Pheobe16t0}.
Constraints on the low-redshift thermal state would thus provide an
important check on our theoretical understanding of the IGM and shed
light on the degree to which any other processes such as blazar
heating, feedback from galaxy formation, or any other exotic physics
can inject heat into the IGM. In an earlier study using the Space Telescope Imaging Spectrograph, 
\citet{Dave01} obtained preliminary evidence that the $T_0$ is indeed about $5000$ K at $z\sim 0$, 
but they also found that the observed low-$z$ Lyman-$\alpha$ lines are not consistent with pure 
thermal broadening and may therefore also be broadened by some additional processes such as some 
type of feedback. Similar issues with line broadening are also reported by 
\citet{Gaikwad17b,Viel17} and \citet{Nasir17}. We can now revisit these issues with much larger 
sample. 

In this paper, we present new measurements of the Lyman-$\alpha$ forest flux power spectrum at 
$z<0.5$ in five redshift bins.  
We use high quality Lyman-$\alpha$ forest spectra (S/N per pixel $>10$) 
observed in 65 background quasars
from the sample of \citet{Danforth14}. 
Combining these power spectrum measurements with state-of-the-art cosmological hydrodynamical 
simulations run with the Nyx code \citep{Almgren13,Lukic15}, we constrain the intensity of UV 
background \(\Gamma_\mathrm{HI}\) at $z<0.5$. 
Our UV background measurements are consistent with recent studies \citep{Shull15, Gaikwad17a, 
Gaikwad17b,Fumagalli17} which confirm that there is no crisis with UV photon production at $z<0.5$ 
and that the primary contributors to the UV background are quasars. 
In a companion paper (Walther et al. in prep.), we will use these power-spectrum measurements to 
constrain the thermal state of the IGM at $z<0.5$.

The paper is organized as follows. In Section 2, we discuss the Lyman-$\alpha$ forest data. In 
Section 3, we describe our method to compute the power spectrum and present the resulting 
measurements. In Section 4, we discuss the implications of our power spectrum regarding the UV 
ionizing background and compare with previous work. In Section 5, we present our conclusions and 
discuss future directions.
Throughout the paper, we adopt a flat $\Lambda$CDM cosmology with parameters $\Omega_m=0.319181$, 
$\Omega_b h^2 = 0.022312$, $h = 0.670386$, $n_s = 0.96$, and $\sigma_8 = 0.8288$ consistent with 
\citet{Planck18}. This cosmology is used
both for our power spectrum measurements as well as for our
cosmological hydrodynamical simulations. All of the distances quoted are comoving.

\begin{figure*}
\includegraphics[width=\textwidth,height=\textheight,keepaspectratio]{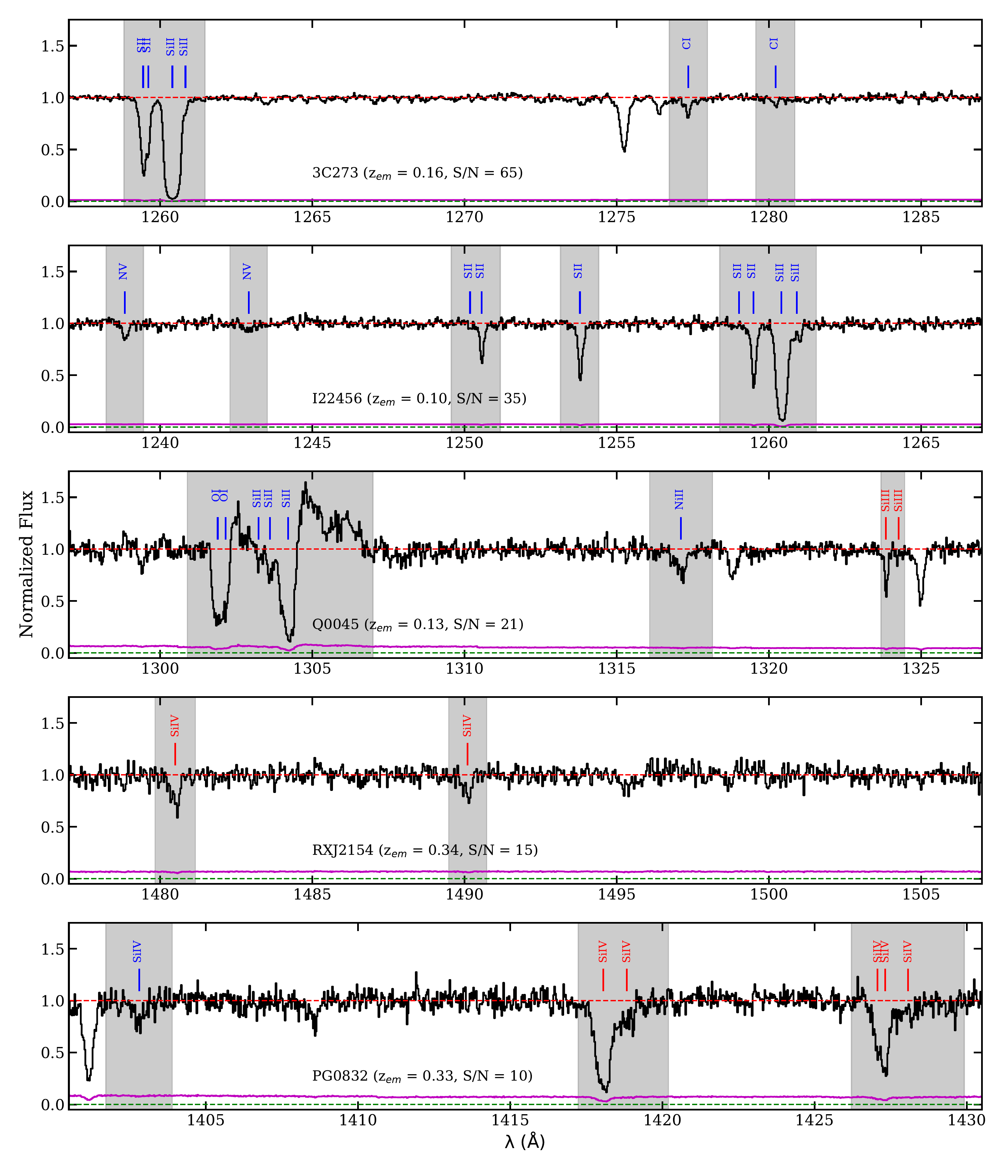}
\caption{Illustration of our masking procedure. Different panels show random 30 \AA~regions of 
five quasar spectra with different S/N per pixel as denoted in panels, in decreasing order from 
top to bottom panel. The red and green dash lines indicate continuum ($y=1$ line) and zero level 
($y=0$ line). Magenta curve shows the error in the normalized flux. The shaded regions show our 
masks. Blue and red ticks indicate the metal absorption from the ISM of the Milky Way and from 
intervening absorbers, respectively. We also mask geocoronal airglow emissions as shown in middle 
panel associated with O {\sc i} ($\lambda \sim 1305$ \AA).}
\label{fig1}
\end{figure*}

\section{Data and Masking} \label{sec:data}
\begin{figure*}
\includegraphics[width=\textwidth,height=\textheight,keepaspectratio]{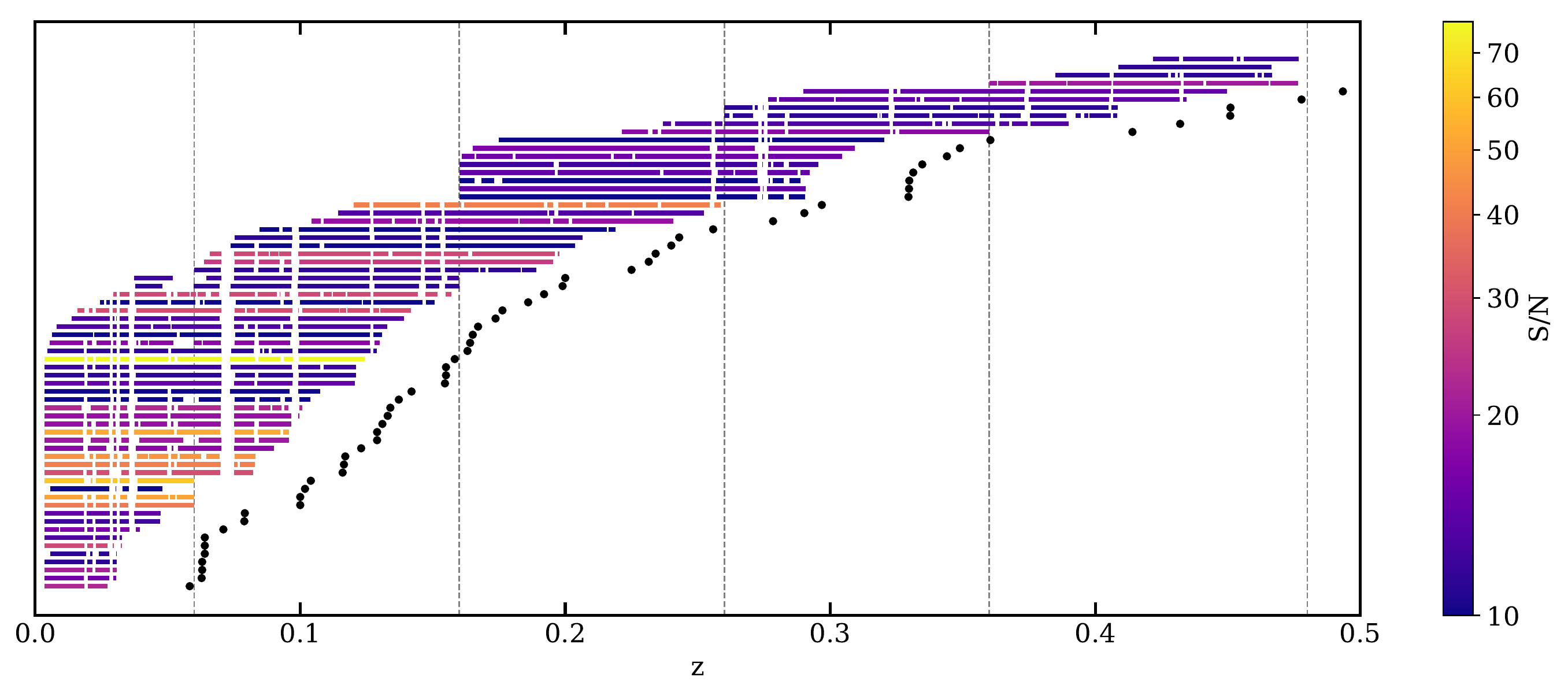}
\caption{
Redshift path covered by the Lyman-$\alpha$ forest used for our power-spectrum calculations. 
Horizontal lines indicate the Lyman-$\alpha$ redshift path (rest frame 1050 to 1180 \AA), filled 
circles show the emission redshift of quasars,  and vertical dashed lines demarcate the redshift 
bins used (i.e, 0.005-0.06, 0.06-0.16, 0.16-0.26, 0.26-0.36 and 0.36-0.48). The gaps in the horizontal 
lines are the masked regions of spectra (see e.g Fig.~\ref{fig1}) containing metal lines, emission 
lines, spectral-gaps, and bad data. This is a subsample of spectra from \citet{Danforth14} where 
the the unmasked Lyman-$\alpha$ forest has S/N (per COS pixel) $\ge 10$, as indicated with a 
colorbar.}
\label{fig2}
\end{figure*}

We use high-quality
medium-resolution ($R\sim 18000$, $\Delta v \sim 17$ km s$^{-1}$)
quasar spectra obtained from HST/COS as a part of the large low-$z$ IGM survey by 
\citet{Danforth14}. This survey contains 82 quasar spectra at $z_{\rm em}<0.72$ observed in the 
wavelength range from $1130-1800$ \AA, which covers the Lyman-$\alpha$ forest at $z<0.48$.
The observations were obtained with the G130M and G160M gratings between year 2009 and 2013. 
\citet{Danforth14} co-added individual spectra (combining both gratings whenever available), 
fitted continua, and identified nearly all individual absorption and emission lines. 
We use these continuum-fitted spectra along with the line catalog made publicly available by 
\citet{Danforth14} as a high-level science product at Mikulski Archive for Space
Telescopes\footnote{Link: http://archive.stsci.edu/prepds/igm/}. 

To calculate the Lyman-$\alpha$ forest flux power spectrum, P$(k)$,
we mask all absorption lines arising from higher Lyman series transitions than Lyman-$\alpha$, the 
metal lines arising from intervening systems and the interstellar medium (ISM) of the Milky Way, 
all emission lines including geocoronal airglow emission, low quality data having S/N $<5$ per 
pixel\footnote{A pixel in this dataset corresponds to $\Delta v = 6.67$ km s$^{-1}$.}, and all gaps
in the spectral coverage. To illustrate our masking procedure, we show five random chunks of 
spectra with different S/N in Fig.~\ref{fig1}. The shaded regions in Fig.~\ref{fig1} show our 
masks. In the Lyman-$\alpha$ forest, masking is critical to removing metal contamination and 
estimating the power spectrum correctly \citep[see, e.g,][]{Walther18a}.
We discuss the effect of not masking metals on the power spectrum in Appendix~\ref{A}. 

After masking, we restrict our further analysis to the rest-frame wavelength range between 1050 and
1180 \AA~in each quasar spectrum to avoid proximity zones of quasars ($\lambda>1180$ \AA) and the 
excessive masking of data due to the presence of higher Lyman series forest lines 
($\lambda<1050$\AA). The redshift path covered by the unmasked Lyman-$\alpha$ forest data is shown 
in Fig.~\ref{fig2} where the quasars are ordered vertically by increasing redshift.
Gaps in the horizontal lines show our masking. Vertically aligned gaps indicate masking due to 
strong Milky Way metal lines. A large gap at 1305 \AA~($z\sim 0.07$) is due to a combination of an 
O~{\sc i} geocoronal airglow emission line and strong Si~{\sc ii} and O~{\sc i} absorption lines 
from the Milky Way (see e.g the middle panel of Fig.~\ref{fig1}).

After masking and choosing the relevant wavelength range, we calculate the median S/N per pixel in 
the unmasked regions and impose a median S/N $>10$ (per pixel) cut to the 82 quasar spectra. 
Although in our power spectrum calculation we subtract the noise (see Section 3), we choose this 
S/N cut to ensure that we are not sensitive to systematic errors associated with how well we know 
the properties of the noise. 
Note that our S/N estimate is different than the default values provided in \citet{Danforth14} who 
computed the S/N per resolution element over the entire spectrum. After applying our S/N cut, we 
are left with the Lyman-$\alpha$
forest of 66 quasars out of the initial 82. Individual values of this S/N per pixel are indicated 
in Fig.~\ref{fig2} via different colors.

We then split the total redshift path covered by these 66 high S/N Lyman-$\alpha$ forest spectra 
into five redshift bins as illustrated by the vertical dashed lines in Fig.~\ref{fig2} and 
summarized in  Table~\ref{tab1}. 
The first bin is chosen from
$z=0.005$ to $0.06$ to remove any systematics arising from the extended wings of geocoronal 
Lyman-$\alpha$ emission line, which sets the lower limit of this redshift bin.
The next three bins ($z=0.06-0.16$, $z=0.16-0.26$ and $z=0.26-0.36$) are chosen to have the same 
width ($\Delta z =0.1$) and also because the mean redshifts of the Lyman-$\alpha$ forests in these 
bins are nicely centered at $z=0.1, \,0.2$, and $0.3$ where we can compare our UV background 
measurements with previous studies \citep[e.g][]{Shull15, Gaikwad17a, Gaikwad17b}. 
The last redshift bin ($z=0.36-0.48$) encloses the remaining redshift-path covered by the data. 
Finally, in each redshift bin we removed short spectra that span less than 10\% of the redshift 
bin-width. This criterion removes one more spectrum from the last bin and we are left with a total 
of 65 quasar spectra shown in Fig.~\ref{fig2}.

\renewcommand{\arraystretch}{1.2}
\begin{table}
\centering
\caption{ Details of the data and comparison simulation}
\begin{tabular}{ccccc}
\hline
Redshift      & ${\bar z}^a$ &Number & Simulation & Simulation \\
bin    & & of quasars &  redshift & T$_0$ (K) and  $\gamma$\\ \hline
 0.005 - 0.06 & 0.03  & 39 &  0.03  & 5033\,\, 1.73\\
 0.06 - 0.16  & 0.10  & 34 &  0.10  & 5288\,\, 1.72\\ 
 0.16 - 0.26  & 0.20  & 19 &  0.20  & 5652\,\, 1.71\\ 
 0.26 - 0.36  & 0.30  & 13 &  0.30  & 6010\,\, 1.69\\ 
 0.36 - 0.48  & 0.41  & 9  &  0.40  & 6368\,\, 1.68\\ 
 \hline
\end{tabular}
\begin{flushleft}
\footnotesize{$^a$Mean redshift of the unmasked Lyman-$\alpha$ forest in the bin.} \\
\end{flushleft}

\label{tab1}
\end{table}

\section{Power Spectrum} \label{sec:ps}

In this section we discuss our method for measuring the power spectrum and present the measurement.

\subsection{Method}\label{sec3.1}
Once the data are prepared, we calculate the P($k$) following the method presented in 
\citet{Walther18a}.  A brief description of the method is as follows. We first calculate the flux 
contrast of each spectrum in a redshift bin $\delta_F = (F- \bar{F})/\bar{F} $ where $\bar{F}$ is 
the mean flux of that spectral chunk in the bin.
Then we use a Lomb-Scargle periodogram \citep{Lomb76,Scargle82} to calculate the raw power spectrum
$P_{\rm raw}(k)$. We subtract off the noise power $P_{\rm noise}(k)$ from $P_{\rm raw}(k)$ and 
divide the difference by the square of the window function $W(k, R)$ corresponding to the 
appropriate COS line spread function (LSF) to correct for finite resolution and pixelization 
\citep[for more details see][]{Palanque13ps, Walther18a}. 
Therefore, our final power spectrum is
\begin{equation}\label{eq1}
P(k)=\bigg \langle \, \frac{P_{\rm raw}(k)-P_{\rm noise}(k) }{W^2(k, R)} \, \bigg \rangle .
\end{equation}
We use the same logarithmic binning used in \citet{Walther18a}, and the average is performed over 
individual Lomb-Scargle periodograms of all Lyman-$\alpha$ forest chunks in each bin.
We follow the standard normalization of the power spectrum, i.e. the variance in the flux contrast 
is $\sigma^2_{\delta_F}=\int_{-\infty}^{\infty}{dk \, P(k)/2\pi}$. 
The noise power within a bin is calculated using many realizations of Gaussian random 
noise\footnote{Since our S/N cut is $\ge 10$, we always record sufficient photons to be in the 
Gaussian regime.}
generated from the error vector of each spectrum of the bin. 
For estimating the window function, we used the COS LSFs corresponding to the different gratings 
and lifetime positions\footnote{We use the python package {\tt linetools} 
(\url{https://linetools.readthedocs.io/en/latest/api.html}) 
which can interpolate the COS LSFs from 
\url{http://www.stsci.edu/hst/cos/performance/spectral_resolution/} 
to any central wavelength.} depending on the observational parameters for each spectrum. 
In particular, spectra at $\bar{z}=0.2$ bin have overlapping contribution from both the
G130M and the 160M grating. In this redshift bin,
motivated by the co-addition routine used by \citet{Danforth14} we take 1460 \AA~as the wavelength 
where the transition between the gratings happen.\footnote{Using a different wavelength in the 
range $1400-1500$ \AA~for the transition from G130M to G160M grating has negligible effect on the 
calculated power spectrum at $\bar{z}=0.2$.}

The COS LSF is quite different from the typical Gaussian LSFs that govern ground based 
spectroscopic observations and exhibits broad wings. In Appendix~\ref{C},
we discuss the effect of incorrectly assuming a Gaussian LSF on the obtained $P(k)$ instead of 
using the correct non-Gaussian COS LSF. Finally, we calculate 
the uncertainties in P$(k)$, the diagonal elements of covariance matrix $C_{ii}\equiv\sigma_i^2$, 
by bootstrap resampling using $10^4$ random realizations of the dataset. Note that we do not have 
enough data to estimate the full covariance matrix $C_{ij}$. We recommend that researchers 
attempting to fit our power spectrum
calculate the full covariance matrix by computing the correlation matrix
$R_{ij} = C_{ij}\slash \sigma_i \sigma_j$ from their models, which can then be scaled by our 
diagonal elements to determine the full covariance matrix \citep[see e.g.][for 
details]{Walther18a}. Our P($k$) measurements and the diagonal elements of covariance matrix are 
given in Table~\ref{tab2}.

\subsection{Results}
\begin{figure*}
\includegraphics[width=0.995\textwidth,height=\textheight,keepaspectratio]{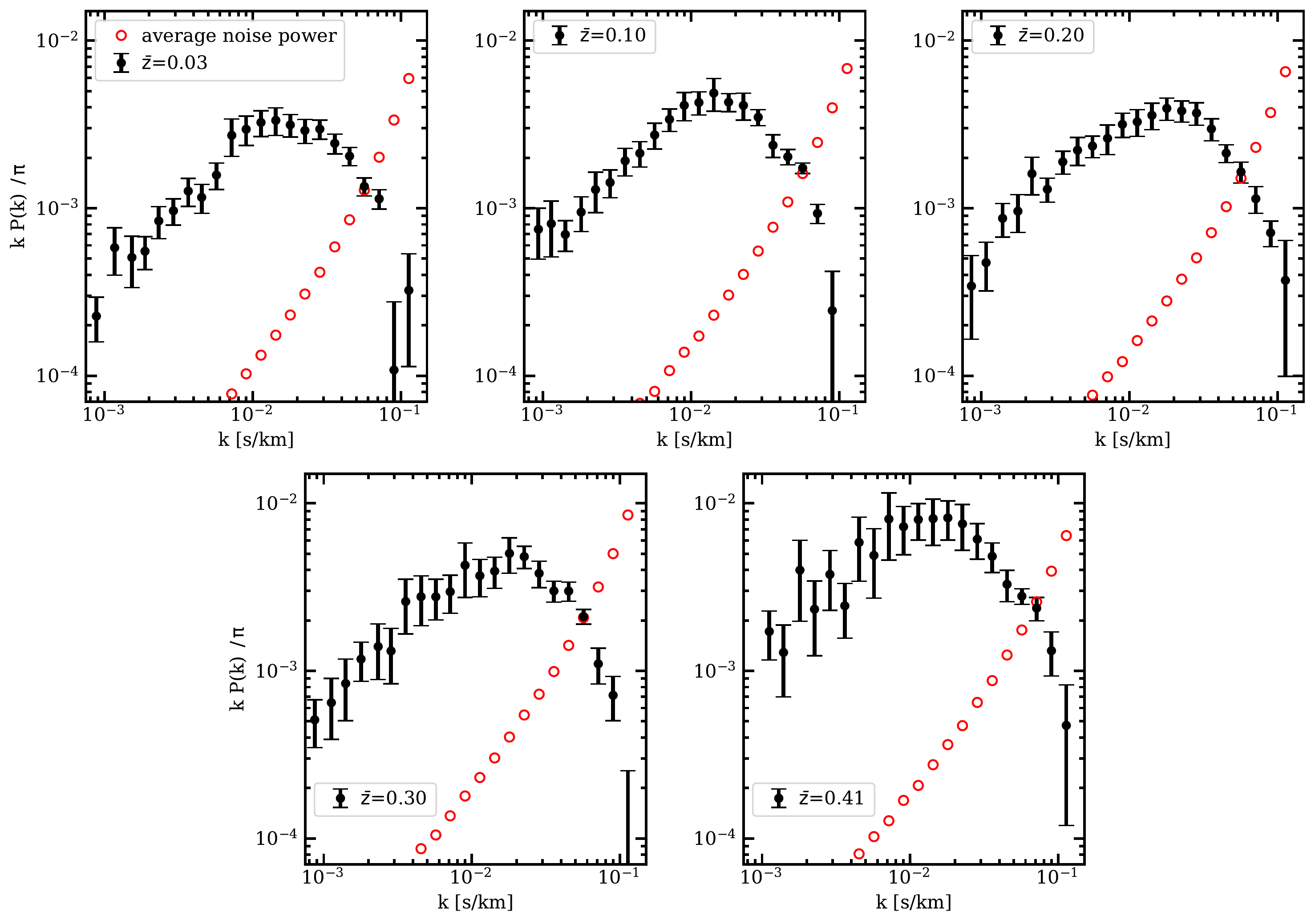}
\caption{ Our Lyman-$\alpha$ forest flux power spectrum ($k$ versus $kP(k)/\pi$) in different 
redshift bins (black points; see Table~\ref{tab1} for more details) at  $k=8\times 10^{-4}$ to 0.15
s km$^{-1}$. The complete measurements are provided in Table~\ref{tab2}. Red points show average 
noise power $\langle \, \frac{P_{\rm noise}(k) }{W^2(k, R)} \rangle$ that we subtract from raw 
power to obtain the power spectrum measurements (see Eq.~\ref{eq1}). The average noise power is 
always less than 5\% of the final power at $k<0.01$ s/km. }  
\label{fig3}
\end{figure*}

In Fig~\ref{fig3}, we show our power spectrum measurement in different redshift bins, which 
reliably probes the power over the range  $k\sim 5\times 10^{-4}$ to 0.15 s km$^{-1}$ (see 
Table~\ref{tab2}). In contrast to $z>2$, the relative ease of identifying and masking  metal 
absorption lines at these low redshifts allows us to probe the small-scale (high-$k$) 
power-spectrum with negligible systematics due to non-Lyman-\(\alpha\) absorption. We discuss the 
effect of not masking metals on the power spectrum in Appendix~\ref{A}.  

At all redshifts, our measured power spectrum shows a clear small scale cut-off at $k>0.03$ s 
km$^{-1}$. This cut-off is a signature of pressure smoothing and Doppler broadening of the 
Lyman-$\alpha$ forest. 
At small scales, the Lyman-$\alpha$ forest is supported by the thermal-pressure and does not 
follow the dark-matter density fluctuations \citep{Hui97,Kulkarni15,Onorbe17,Rorai17,Nasir17}. 
This pressure support, in addition to the Doppler broadening, smooths out the fluctuations in the 
Lyman-$\alpha$ forest flux and gives rise to the small-scale cut-off seen in the power spectrum 
\citep[see e.g,][]{Peeples10,Rorai13}. This cut-off is an important feature that probes the 
thermal state of the IGM \citep{Zaldarriaga01, Walther19}. 

The amplitude of the power spectra at all $k$ is significantly smaller than that obtained at 
high-$z$ \citep[see, e.g,][]{Walther18a}. This is because the density evolution of the Universe 
results in lower overall
opacity in the low-z IGM  giving rise to a thinner low-z Lyman-$\alpha$ forest.
This reduced opacity also reduces the power at all scales. Nevertheless, we have obtained high 
precision measurements (15\% at $z<0.2$, 25\% at $z=0.3$ and 30\% at $z=0.41$ over $0.001<k<0.1$ s
km$^{-1}$ scale) of the power spectrum because of the large sample size. In contrast to high-$z$, 
the redshift evolution of the amplitude of the power appears quite shallow. This
evolution is so weak that it is challenging to identify; for example, our large scale power 
($k<0.02$ s km$^{-1}$) measurement at $\bar z =0.2$ is only slightly lower than that at $\bar 
z=0.1$. Such non-monotonic
redshift evolution is unexpected but likely results from a combination of
noise fluctuations and very shallow redshift evolution. Indeed, extremely shallow evolution in the
power spectrum amplitude is not unexpected at low-$z$. 
The power spectrum $P(k,z)\propto \langle FF^\ast\rangle \propto e^{-2\tau(z)}$ scales with the 
evolution of the optical depth $\tau(z)$. Following the
fluctuating Gunn-Peterson approximation (FGPA), we can write 
\citep[][]{Gunn65,Croft98}
\begin{equation}\label{eq2}
\tau(z) \propto  \, \Gamma_{\rm HI} ^{-1} \, T_0^{-0.7} \, n_{\rm H}^2, 
\end{equation}
where $\Gamma_{\rm HI}$ is the photoionization rate of \HI\ from the UV background (UVB) and 
$n_{\rm H}$ is the hydrogen
density. Using $\Gamma_{\rm HI}(z) \propto (1+z)^5$ \citep[from][]{Shull15, Gaikwad17a}, 
$n_{\rm H}(z) \propto (1+z)^3$ from cosmological density evolution, and assuming a power law 
redshift evolution for
$T_0(z) \propto (1+z)^{\beta}$, the opacity of the IGM should scale as $\tau \propto 
(1+z)^{1-0.7{\beta}}$. In
the absence of any non-standard heating processes, theory predicts a
cool-down of the IGM
at low redshift suggesting $\beta>0$ \citep[see e.g,][]{Pheobe16t0}.
This suggests that $\tau(z)$ and the
resulting amplitude of the power spectrum evolve slowly at low redshifts. However, note that the 
FGPA is not a good approximation for the low-$z$ IGM, at least for the power spectrum at large 
scales (small $k$), because
it does not include the effects of shock-heated gas, 
as explained in Section 4 and Fig.~\ref{fig4}. Therefore, the redshift evolution of the amplitude 
of the large scale power is likely to be more complicated than the simple picture presented here.

The flux power spectrum at redshifts $0.1\le z \le 0.4$ was also presented by \citet{Gaikwad17a}, 
however they split observed spectra into chunks of size 50 cMpc/h to compare with their simulation
box size for the specific purpose of only evaluating the UV background. 
Also their method of calculating the power spectrum and the normalization is quite different from 
ours. For example, they fill the masks with added continuum and random noise and estimate the 
power spectrum of the flux ($F$) rather than the flux-contrast ($\delta F$). 
For these reasons, we are unable to compare with the \citet{Gaikwad17a} power spectrum 
measurements directly. However, we can compare with their UV background measurements, which use 
not only the power-spectrum but also the flux probability density function (PDF) and 
column-density distribution function \citep[CDDF;][]{Gaikwad17b}. 

\begin{table*}
\let\clearpage\relax
\centering
\def\arraystretch{0.8}%
\caption{Power spectrum measurement}
\begin{tabular}{cccc|cccc}
\hline
${\bar z}$& $k (s/km)$ & $kP(k)/\pi$ &  $\sigma_{kP(k)/\pi}$& ${\bar z}$& $k (s/km)$ & $kP(k)/\pi$ &  $\sigma_{kP(k)/\pi}$\\
\hline
0.03 &3.937$\times 10^{-4}$ &1.874$\times 10^{-4}$ &7.550$\times 10^{-5}$&0.20 &2.794$\times 10^{-3}$ &1.298$\times 10^{-3}$ &2.129$\times 10^{-4}$ \\
0.03 &4.390$\times 10^{-4}$ &1.228$\times 10^{-4}$ &5.164$\times 10^{-5}$&0.20 &3.551$\times 10^{-3}$ &1.892$\times 10^{-3}$ &2.949$\times 10^{-4}$\\
0.03 &7.825$\times 10^{-4}$ &3.137$\times 10^{-4}$ &9.136$\times 10^{-5}$&0.20 &4.466$\times 10^{-3}$ &2.219$\times 10^{-3}$ &4.231$\times 10^{-4}$\\
0.03 &8.805$\times 10^{-4}$ &2.267$\times 10^{-4}$ &6.759$\times 10^{-5}$&0.20 &5.636$\times 10^{-3}$ &2.348$\times 10^{-3}$ &3.451$\times 10^{-4}$ \\
0.03 &1.170$\times 10^{-3}$ &5.812$\times 10^{-4}$ &1.833$\times 10^{-4}$&0.20 &7.117$\times 10^{-3}$ &2.612$\times 10^{-3}$ &5.248$\times 10^{-4}$ \\
0.03 &1.528$\times 10^{-3}$ &5.083$\times 10^{-4}$ &1.726$\times 10^{-4}$&0.20 &8.969$\times 10^{-3}$ &3.162$\times 10^{-3}$ &5.229$\times 10^{-4}$ \\
0.03 &1.874$\times 10^{-3}$ &5.534$\times 10^{-4}$ &1.232$\times 10^{-4}$&0.20 &1.130$\times 10^{-2}$ &3.278$\times 10^{-3}$ &5.957$\times 10^{-4}$ \\
0.03 &2.322$\times 10^{-3}$ &8.393$\times 10^{-4}$ &1.814$\times 10^{-4}$&0.20 &1.420$\times 10^{-2}$ &3.584$\times 10^{-3}$ &6.451$\times 10^{-4}$ \\
0.03 &2.915$\times 10^{-3}$ &9.651$\times 10^{-4}$ &1.717$\times 10^{-4}$&0.20 &1.788$\times 10^{-2}$ &3.943$\times 10^{-3}$ &5.999$\times 10^{-4}$ \\
0.03 &3.670$\times 10^{-3}$ &1.265$\times 10^{-3}$ &2.399$\times 10^{-4}$&0.20 &2.255$\times 10^{-2}$ &3.809$\times 10^{-3}$ &5.501$\times 10^{-4}$ \\
0.03 &4.523$\times 10^{-3}$ &1.161$\times 10^{-3}$ &2.277$\times 10^{-4}$&0.20 &2.838$\times 10^{-2}$ &3.695$\times 10^{-3}$ &5.647$\times 10^{-4}$ \\
0.03 &5.691$\times 10^{-3}$ &1.576$\times 10^{-3}$ &2.837$\times 10^{-4}$&0.20 &3.576$\times 10^{-2}$ &2.974$\times 10^{-3}$ &4.456$\times 10^{-4}$ \\
0.03 &7.217$\times 10^{-3}$ &2.721$\times 10^{-3}$ &6.823$\times 10^{-4}$&0.20 &4.503$\times 10^{-2}$ &2.129$\times 10^{-3}$ &2.573$\times 10^{-4}$ \\
0.03 &9.028$\times 10^{-3}$ &2.958$\times 10^{-3}$ &5.895$\times 10^{-4}$&0.20 &5.667$\times 10^{-2}$ &1.648$\times 10^{-3}$ &2.354$\times 10^{-4}$ \\
0.03 &1.137$\times 10^{-2}$ &3.248$\times 10^{-3}$ &5.704$\times 10^{-4}$&0.20 &7.131$\times 10^{-2}$ &1.138$\times 10^{-3}$ &2.070$\times 10^{-4}$ \\
0.03 &1.431$\times 10^{-2}$ &3.343$\times 10^{-3}$ &6.273$\times 10^{-4}$&0.20 &8.975$\times 10^{-2}$ &7.136$\times 10^{-4}$ &1.244$\times 10^{-4}$ \\
0.03 &1.792$\times 10^{-2}$ &3.142$\times 10^{-3}$ &4.820$\times 10^{-4}$&0.20 &1.130$\times 10^{-1}$ &3.709$\times 10^{-4}$ &2.719$\times 10^{-4}$ \\
0.03 &2.250$\times 10^{-2}$ &2.907$\times 10^{-3}$ &4.776$\times 10^{-4}$&0.30 &5.576$\times 10^{-4}$ &4.136$\times 10^{-4}$ &2.045$\times 10^{-4}$ \\
0.03 &2.837$\times 10^{-2}$ &2.966$\times 10^{-3}$ &3.903$\times 10^{-4}$&0.30 &8.690$\times 10^{-4}$ &5.103$\times 10^{-4}$ &1.624$\times 10^{-4}$ \\
0.03 &3.578$\times 10^{-2}$ &2.438$\times 10^{-3}$ &3.273$\times 10^{-4}$&0.30 &1.126$\times 10^{-3}$ &6.460$\times 10^{-4}$ &2.556$\times 10^{-4}$ \\
0.03 &4.503$\times 10^{-2}$ &2.045$\times 10^{-3}$ &2.542$\times 10^{-4}$&0.30 &1.406$\times 10^{-3}$ &8.403$\times 10^{-4}$ &3.366$\times 10^{-4}$ \\
0.03 &5.666$\times 10^{-2}$ &1.349$\times 10^{-3}$ &1.670$\times 10^{-4}$&0.30 &1.786$\times 10^{-3}$ &1.175$\times 10^{-3}$ &3.081$\times 10^{-4}$ \\
0.03 &7.133$\times 10^{-2}$ &1.137$\times 10^{-3}$ &1.502$\times 10^{-4}$&0.30 &2.330$\times 10^{-3}$ &1.396$\times 10^{-3}$ &5.073$\times 10^{-4}$ \\
0.03 &8.982$\times 10^{-2}$ &1.082$\times 10^{-4}$ &1.681$\times 10^{-4}$&0.30 &2.846$\times 10^{-3}$ &1.315$\times 10^{-3}$ &4.777$\times 10^{-4}$ \\
0.03 &1.131$\times 10^{-1}$ &3.235$\times 10^{-4}$ &2.101$\times 10^{-4}$&0.30 &3.575$\times 10^{-3}$ &2.593$\times 10^{-3}$ &9.301$\times 10^{-4}$ \\
0.10 &2.397$\times 10^{-4}$ &2.721$\times 10^{-4}$ &1.521$\times 10^{-4}$&0.30 &4.537$\times 10^{-3}$ &2.771$\times 10^{-3}$ &9.136$\times 10^{-4}$\\  
0.10 &2.861$\times 10^{-4}$ &1.436$\times 10^{-4}$ &4.891$\times 10^{-5}$&0.30 &5.706$\times 10^{-3}$ &2.767$\times 10^{-3}$ &7.532$\times 10^{-4}$\\  
0.10 &3.521$\times 10^{-4}$ &1.093$\times 10^{-4}$ &5.191$\times 10^{-5}$&0.30 &7.137$\times 10^{-3}$ &2.966$\times 10^{-3}$ &7.580$\times 10^{-4}$ \\ 
0.10 &4.724$\times 10^{-4}$ &5.567$\times 10^{-4}$ &4.046$\times 10^{-4}$&0.30 &8.992$\times 10^{-3}$ &4.268$\times 10^{-3}$ &1.529$\times 10^{-3}$ \\ 
0.10 &5.742$\times 10^{-4}$ &3.166$\times 10^{-4}$ &8.959$\times 10^{-5}$&0.30 &1.131$\times 10^{-2}$ &3.695$\times 10^{-3}$ &9.241$\times 10^{-4}$ \\ 
0.10 &7.142$\times 10^{-4}$ &3.239$\times 10^{-4}$ &1.059$\times 10^{-4}$&0.30 &1.424$\times 10^{-2}$ &3.936$\times 10^{-3}$ &8.243$\times 10^{-4}$ \\ 
0.10 &9.321$\times 10^{-4}$ &7.483$\times 10^{-4}$ &2.526$\times 10^{-4}$&0.30 &1.793$\times 10^{-2}$ &5.021$\times 10^{-3}$ &1.199$\times 10^{-3}$ \\ 
0.10 &1.139$\times 10^{-3}$ &8.075$\times 10^{-4}$ &2.949$\times 10^{-4}$&0.30 &2.256$\times 10^{-2}$ &4.804$\times 10^{-3}$ &7.301$\times 10^{-4}$ \\ 
0.10 &1.417$\times 10^{-3}$ &6.968$\times 10^{-4}$ &1.449$\times 10^{-4}$&0.30 &2.842$\times 10^{-2}$ &3.819$\times 10^{-3}$ &6.866$\times 10^{-4}$ \\ 
0.10 &1.812$\times 10^{-3}$ &9.472$\times 10^{-4}$ &2.221$\times 10^{-4}$&0.30 &3.576$\times 10^{-2}$ &2.999$\times 10^{-3}$ &4.241$\times 10^{-4}$ \\ 
0.10 &2.265$\times 10^{-3}$ &1.291$\times 10^{-3}$ &3.512$\times 10^{-4}$&0.30 &4.499$\times 10^{-2}$ &2.994$\times 10^{-3}$ &3.940$\times 10^{-4}$ \\ 
0.10 &2.841$\times 10^{-3}$ &1.423$\times 10^{-3}$ &2.675$\times 10^{-4}$&0.30 &5.664$\times 10^{-2}$ &2.110$\times 10^{-3}$ &2.131$\times 10^{-4}$ \\ 
0.10 &3.590$\times 10^{-3}$ &1.917$\times 10^{-3}$ &3.613$\times 10^{-4}$&0.30 &7.130$\times 10^{-2}$ &1.101$\times 10^{-3}$ &2.666$\times 10^{-4}$ \\ 
0.10 &4.511$\times 10^{-3}$ &2.127$\times 10^{-3}$ &3.669$\times 10^{-4}$&0.30 &8.975$\times 10^{-2}$ &7.149$\times 10^{-4}$ &2.101$\times 10^{-4}$ \\ 
0.10 &5.677$\times 10^{-3}$ &2.735$\times 10^{-3}$ &4.798$\times 10^{-4}$&0.41 &1.118$\times 10^{-3}$ &1.717$\times 10^{-3}$ &5.549$\times 10^{-4}$ \\ 
0.10 &7.151$\times 10^{-3}$ &3.390$\times 10^{-3}$ &5.174$\times 10^{-4}$&0.41 &1.395$\times 10^{-3}$ &1.287$\times 10^{-3}$ &5.887$\times 10^{-4}$ \\ 
0.10 &8.990$\times 10^{-3}$ &4.109$\times 10^{-3}$ &7.895$\times 10^{-4}$&0.41 &1.801$\times 10^{-3}$ &3.991$\times 10^{-3}$ &2.017$\times 10^{-3}$ \\ 
0.10 &1.130$\times 10^{-2}$ &4.269$\times 10^{-3}$ &6.702$\times 10^{-4}$&0.41 &2.259$\times 10^{-3}$ &2.333$\times 10^{-3}$ &1.102$\times 10^{-3}$ \\ 
0.10 &1.423$\times 10^{-2}$ &4.864$\times 10^{-3}$ &1.069$\times 10^{-3}$&0.41 &2.870$\times 10^{-3}$ &3.762$\times 10^{-3}$ &1.463$\times 10^{-3}$ \\ 
0.10 &1.792$\times 10^{-2}$ &4.292$\times 10^{-3}$ &5.291$\times 10^{-4}$&0.41 &3.616$\times 10^{-3}$ &2.445$\times 10^{-3}$ &8.784$\times 10^{-4}$ \\ 
0.10 &2.255$\times 10^{-2}$ &4.104$\times 10^{-3}$ &7.505$\times 10^{-4}$&0.41 &4.516$\times 10^{-3}$ &5.841$\times 10^{-3}$ &2.415$\times 10^{-3}$ \\ 
0.10 &2.839$\times 10^{-2}$ &3.489$\times 10^{-3}$ &3.827$\times 10^{-4}$&0.41 &5.687$\times 10^{-3}$ &4.887$\times 10^{-3}$ &2.170$\times 10^{-3}$ \\ 
0.10 &3.573$\times 10^{-2}$ &2.375$\times 10^{-3}$ &3.648$\times 10^{-4}$&0.41 &7.172$\times 10^{-3}$ &8.052$\times 10^{-3}$ &3.469$\times 10^{-3}$ \\ 
0.10 &4.497$\times 10^{-2}$ &2.027$\times 10^{-3}$ &2.117$\times 10^{-4}$&0.41 &9.012$\times 10^{-3}$ &7.250$\times 10^{-3}$ &2.318$\times 10^{-3}$ \\ 
0.10 &5.663$\times 10^{-2}$ &1.736$\times 10^{-3}$ &1.238$\times 10^{-4}$&0.41 &1.133$\times 10^{-2}$ &7.995$\times 10^{-3}$ &1.968$\times 10^{-3}$ \\ 
0.10 &7.131$\times 10^{-2}$ &9.309$\times 10^{-4}$ &1.217$\times 10^{-4}$&0.41 &1.427$\times 10^{-2}$ &8.106$\times 10^{-3}$ &2.493$\times 10^{-3}$\\  
0.10 &8.977$\times 10^{-2}$ &2.450$\times 10^{-4}$ &1.744$\times 10^{-4}$&0.41 &1.793$\times 10^{-2}$ &8.183$\times 10^{-3}$ &2.148$\times 10^{-3}$\\  
0.20 &2.666$\times 10^{-4}$ &9.025$\times 10^{-5}$ &2.044$\times 10^{-5}$&0.41 &2.252$\times 10^{-2}$ &7.539$\times 10^{-3}$ &2.298$\times 10^{-3}$ \\ 
0.20 &5.359$\times 10^{-4}$ &2.741$\times 10^{-4}$ &8.676$\times 10^{-5}$&0.41 &2.835$\times 10^{-2}$ &6.100$\times 10^{-3}$ &1.462$\times 10^{-3}$ \\ 
0.20 &7.364$\times 10^{-4}$ &1.893$\times 10^{-4}$ &5.990$\times 10^{-5}$&0.41 &3.576$\times 10^{-2}$ &4.826$\times 10^{-3}$ &9.621$\times 10^{-4}$ \\ 
0.20 &8.586$\times 10^{-4}$ &3.430$\times 10^{-4}$ &1.778$\times 10^{-4}$&0.41 &4.502$\times 10^{-2}$ &3.285$\times 10^{-3}$ &6.987$\times 10^{-4}$ \\ 
0.20 &1.078$\times 10^{-3}$ &4.733$\times 10^{-4}$ &1.528$\times 10^{-4}$&0.41 &5.666$\times 10^{-2}$ &2.789$\times 10^{-3}$ &2.951$\times 10^{-4}$ \\ 
0.20 &1.394$\times 10^{-3}$ &8.696$\times 10^{-4}$ &1.969$\times 10^{-4}$&0.41 &7.131$\times 10^{-2}$ &2.367$\times 10^{-3}$ &3.749$\times 10^{-4}$ \\ 
0.20 &1.767$\times 10^{-3}$ &9.598$\times 10^{-4}$ &2.439$\times 10^{-4}$&0.41 &8.977$\times 10^{-2}$ &1.319$\times 10^{-3}$ &3.876$\times 10^{-4}$ \\ 
0.20 &2.200$\times 10^{-3}$ &1.607$\times 10^{-3}$ &4.047$\times 10^{-4}$&0.41 &1.131$\times 10^{-1}$ &4.726$\times 10^{-4}$ &3.530$\times 10^{-4}$ \\ 
\hline
\end{tabular}
\begin{flushleft}
\footnotesize{\hspace{70pt} A machine readable version is available on ArXiv ancillary files and on MNRAS webpage.} \\
\end{flushleft}
\label{tab2}
\end{table*}

\section{Implications for the UV background} \label{sec:uvb}

The amplitude of the Lyman-$\alpha$ forest power spectrum is sensitive to the UV background; 
therefore, it can be used to measure the UV background quantified by the \HI~photoionization rate 
$\Gamma_{\mathHI}$. The basic idea behind the measurement is to compare the power-spectrum with 
cosmological hydrodynamical simulations of the IGM where $\Gamma_{\mathHI}$ is one of the free 
parameters. In this section, we first discuss our IGM simulations and then the $\Gamma_{\mathHI}$ 
measurements obtained from them.

\subsection{Simulations}

For comparing our power spectrum measurements with the simulated IGM, we ran an Nyx
cosmological hydrodynamic simulation \citep{Almgren13, Lukic15} from $z=159$ to $z=0.03$.
This is an Eulerian hydrodynamical simulation of box size 20 cMpc/h and 1024$^3$ cells.
The initial conditions were generated using MUSIC code \citep{Hahn11} along with the transfer 
function from CAMB \citep{Lewis00,Howlett12}.
We evolve baryon hydrodynamics in Eulerian approach on a fixed Cartesian grid with 1024$^3$ cells,
and follow the evolution of dark matter using 1024$^3$ Lagrangian (N-body) particles. 
This simulation resolves 19.5 ckpc/h scales (i.e, $\Delta v <2$ km s$^{-1}$ at $z<0.5$). 
In this simulation, we used the photoheating rates from the \citet{Puchwein19} non-equilibrium 
models (the equivalent-equilibrium rates  since our code assumes ionization equilibrium). 
We stored simulation outputs at different redshifts corresponding to our $\bar z$ at the measured 
P($k$). We determined  T$_0$ and  $\gamma$ by  fitting  the  distribution  of  densities and 
temperatures in the simulation following the linear least squares method described in 
\citet{Lukic15}. The simulation redshifts, T$_0$, and $\gamma$ are provided in Table~\ref{tab1}. 
The values of T$_0$ and $\gamma$ are also consistent with the theoretical models presented in 
\citet{McQuinn16} and obtained in the simulations presented by \citet{Shull15} and 
\citet{Gaikwad17a}.

For our current purposes, the only free parameter in this simulation
is $\Gamma_{\rm H\, I}$. We vary $\Gamma_{\rm H\, I}$ and generate the
simulated Lyman-$\alpha$ forest as follows.  First, we calculate the
ionization fractions of hydrogen and helium under the assumption of
ionization equilibrium including both photoionization and collisional
ionization. For this we have used updated cross-sections and
recombination rates from \citet{Lukic15}.  Then, we extract a large
number ($5\times 10^4$) of random lines-of-sight (skewers) parallel to
the (arbitrarily chosen) $z-$axis of the simulation cube.  Along these
lines-of-sight, we store the ionization densities,  temperatures and
$z$-component of the velocities.
As our procedure does not include
radiative transfer, we model the self-shielding of dense cells using
the prescription described in \citet{Rahmati13}.  We generate the
simulated Lyman-$\alpha$ optical depth for each cell along the line of
sight, which we will refer to as $\tau$-skewers, by summing all the
real space contributions to the redshift space optical depth using the
full Voigt profile resulting from each (real-space) cell following the
approximations used in \cite{Tepper06}.  The flux $F=e^{-\tau}$ gives
us the continuum normalized Lyman-$\alpha$ forest flux along these
skewers. These constitute our `perfect skewers' from the
simulation. We calculate and store 5$\times$10$^4$ simulated skewers
from each box for different values of $\Gamma_{\rm H\, I}$.

\begin{figure}
\includegraphics[width=0.48\textwidth,height=\textheight,keepaspectratio]{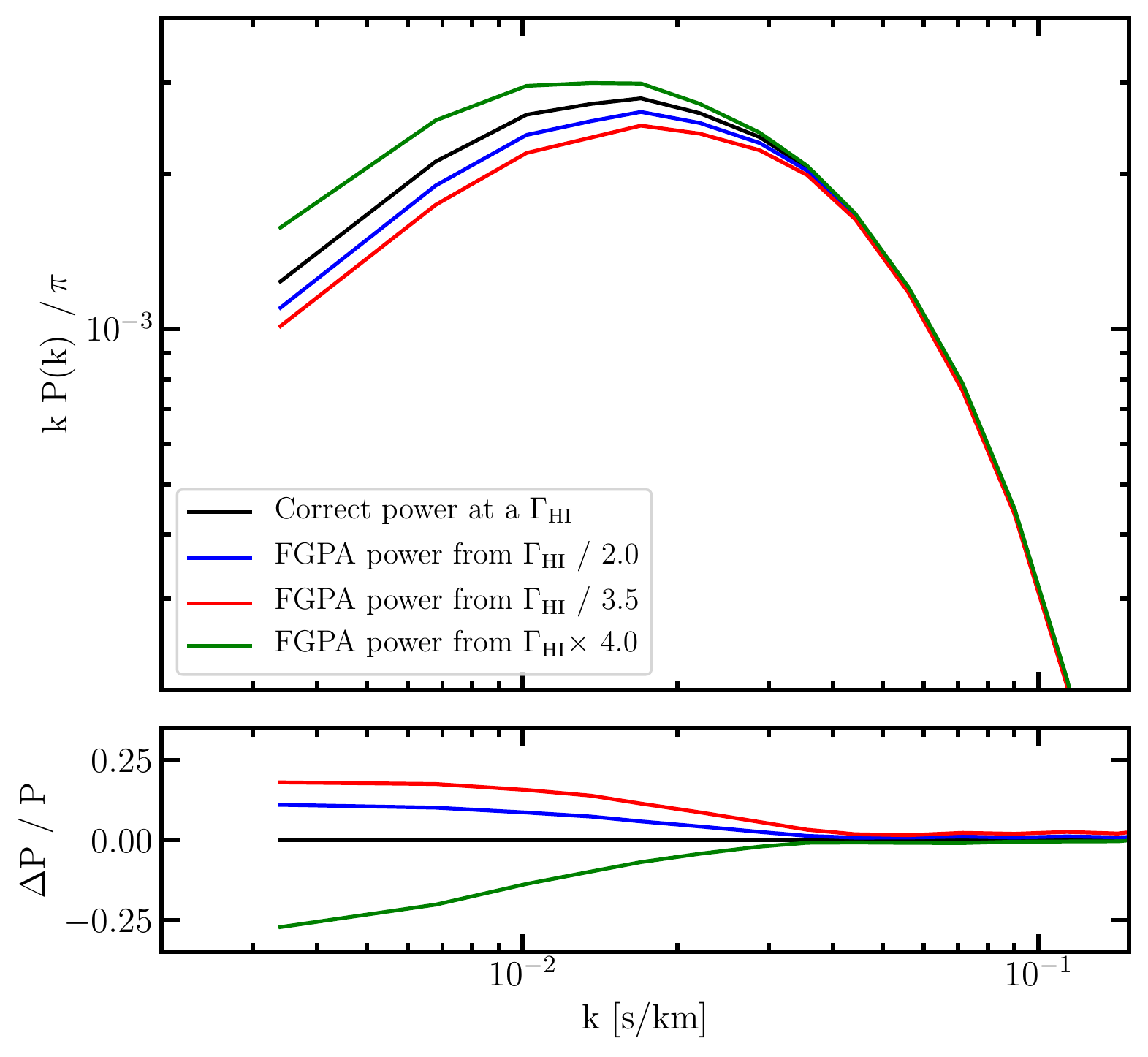}
\caption{The effect of using the FGPA on power-spectra. 
Top panel shows $z=0.2$ power-spectrum estimated for 
$\Gamma_{\rm H\, I\, Fid}=1.75 \times 10^{-13}$ s$^{-1}$ 
(black-curve) by correctly calculating the $\tau$-skewers. 
The green, blue and red curves show power-spectra estimated 
when $\tau$-skewers, which were initially calculated for 
$\Gamma_{\rm H\, I\, Fid} \times 4$,  $\Gamma_{\rm H\, I} /2$ 
and $\Gamma_{\rm H\, I\, Fid} /3.5$ values, were rescaled to 
$\Gamma_{\rm H\, I\, Fid}$ values by following FGPA. At 
$k<0.04$ s km$^{-1}$ the power-spectra obtained using FGPA 
deviates from the correct power-spectrum (black-curve) and 
the percentage deviation are shown in the bottom panel 
with the same colors. The deviation is large for 
large differences in the initial and final $\Gamma_{\rm H\, I}$ values.}
\label{fig4}
\end{figure}

Note that every-time we change the $\Gamma_{\rm H\, I}$ we recalculate
the skewers following the procedure described above.  We do not simply
rescale the $\tau$ values along the skewers when we change the
$\Gamma_{\rm H\, I}$, as is typically applied to simulations of the
Lyman-$\alpha$ forest at higher redshifts following the FGPA.  According
to the FGPA, $\tau \propto \Gamma_{\rm H\, I}^{-1}$ (from Eq.~\ref{eq2}),
which is a good approximation at high-$z$ when most of the gas in the
IGM is photoionized.  However, a large amount of gas in the low-$z$
Universe is collisionally ionized because it is heated to $T>
10^5\,{\rm K}$ by structure formation shocks \citep[see][]{Dave10,
Shull12}. In this regime, the simple rescaling of $\tau$ following the
FGPA leads to erroneous results, because the contribution of collisional ionization
implies that the true optical depth is no longer linearly proportional
to $\Gamma_{\rm H\, I}$ \citep[see also][]{Lukic15}.
This is illustrated in Fig.~\ref{fig4} where we show the
power-spectrum
for $\tau$-skewers generated with a fiducial value of 
$\Gamma_{\rm H\,I \, Fid}=1.75 \times 10^{-13}$ s$^{-1}$ (black-curve) and three
more power-spectra where the $\tau$-skewers were initially calculated
for $\Gamma_{\rm H\, I \, Fid} \times 4$, $\Gamma_{\rm H\, I\, Fid}
/3.5$ and $\Gamma_{\rm H\, I, Fid} /2$ values and then rescaled to
get the $\tau$-skewers corresponding to $\Gamma_{\rm H\, I \, Fid}$
following FGPA.  The latter three deviate significantly 
from the
fiducial power-spectrum on large scales (low $k <0.04$ s km$^{-1}$values). The bottom panel of 
Fig.~\ref{fig4} shows the percentage differences in
the correct calculation versus the ones obtained by using the FGPA.
The differences are large (of the order of 10 to 25\%) when the FGPA is applied for larger
difference in $\Gamma_{\rm H\, I}$ (of factor 2 to 4). Given that the reported low-$z$
$\Gamma_{\rm H\, I}$ measurements vary over factors of $2-5$, the
results obtained by incorrectly using FGPA can give large systematic
differences in the derived $\Gamma_{\rm H\, I}$ values.

Once the perfect-skewers are constructed, we adopt a forward modeling approach to make them look 
like realistic spectra. To this end, we follow the procedure developed in \citet{Walther18a}
but for the HST COS data.  We first stitch randomly drawn skewers
together to cover the Lyman-$\alpha$ forest redshift path of each
quasar. Then, we convolve these skewers with the finite COS LSFs
corresponding to the same gratings and lifetime positions as
the data for each quasar.  Then, we rebin these to match the pixels
of the actual data, and we add Gaussian random noise at each
pixel generated with the standard deviation of the error-vector from
the observed spectra.  Finally, we mask these spectra in exactly same
manner as the data and follow the same procedure to calculate the
power-spectrum (see Section 3.1).  We calculate $P(k)$ from these
forward models created for a large number of $\Gamma_{\rm H\, I}$ values
and estimate the $\Gamma_{\rm H\, I}$ by comparing with the $P(k)$
measurements as explained in the next sub-section.

\subsection{Constraints on the UV background}\label{sec4.2}
The power spectrum is not only sensitive to  $\Gamma_{\rm H\, I}$ 
but also to the thermal state of the IGM quantified by $T_0$ and $\gamma$. 
Therefore, to correctly measure the $\Gamma_{\rm H\, I}$ using our power 
spectrum, we need a large ensemble of simulations of different IGM thermal state models 
stored at the redshifts where we have measured the power spectrum. We have the THERMAL 
grid\footnote{Link: http://thermal.joseonorbe.com/} \citep{Hiss18,Walther19} available 
with $>70$ Nyx simulations (with box size 20 cMpc/h and 1024$^3$ particles) 
of different IGM thermal models but only a single redshift $z=0.2$ overlaps with our dataset 
presented here.
We use 50 simulations from this suite and  follow the Bayesian inference
approach presented in \citet{Walther19}, where the likelihood of the model is given by 
\begin{align}
\mathcal{L}\equiv&P(\mathrm{data}|\mathrm{model})\\
\propto&\prod_\mathrm{datasets}\frac{1}{\sqrt{\det(C)}}\exp\left(-\frac{\mathbf{\Delta}^\mathrm{T} C^{-1} \mathbf{\Delta}}{2}\right) \nonumber\\
\mathbf{\Delta}=&\mathbf{P}_\mathrm{data}-\mathbf{P}_\mathrm{model} \nonumber
\label{eq:likelihood}
\end{align}
Here $C$ is the covariance matrix of the measurements where the diagonal elements are taken from 
the uncertainties obtained here (see Table~\ref{tab2}), and the off-diagonal elements are 
estimated from the forward model which is closest in the parameter space to the model in question.  
Using this $\mathcal{L}$,
we perform a Bayesian inference at $z=0.2$ using Markov Chain Monte Carlo methods (MCMC)
and jointly estimate the $T_0$, $\gamma$, and $\Gamma_{\rm H\, I}$ at $z=0.2$ (Walther et al. in 
prep.). We then marginalize the joint posterior distribution over parameters $T_0$ and $\gamma$ to
determine $\Gamma_{\rm H\, I}$ and its  68\% confidence interval.

At other redshifts, we do not yet  have such a large simulation grid. However the analysis at 
$z=0.2$ provides a clear indication of how much the degeneracies in the thermal state of the low-z
IGM propagate into uncertainties on the $\Gamma_{\rm H\, I}$ measurements. 
To estimate the  $\Gamma_{\rm H\, I}$ at other redshifts but using only one thermal model
rather than a large grid of thermal models, we simply assume that the scaling between 
uncertainties at these redshifts behave similarly to our $z=0.2$ $\Gamma_{\rm H\, I}$ measurement. 
To quantify the amount by which the uncertainties on the  $\Gamma_{\rm H\, I}$ can be 
underestimated if one uses only a single simulation instead of the full grid, we repeat the 
analysis mentioned above at $z=0.2$ but using only one simulation with $\Gamma_{\rm H\, I}$ being 
the only free parameter. 
In this analysis, we are using the 
same simulation for which we have stored outputs at other redshifts as well (as described in 
Section 3 and Table \ref{tab1}). 
Also, for this $\Gamma_{\rm HI}$ estimate, we opt to use only the diagonal elements of the 
covariance
matrix because the shape of power spectrum 
does not vary widely in this single parameter model where
the IGM thermal state is fixed. We found that fitting with the full covariance led to
spurious bad fits resulting from the rigidity of the model of the power spectrum shape if the
thermal state is fixed.
In this case, where we use only one simulation (and hence one thermal model) and the diagonal 
elements of covariances, we calculate the $\Gamma_{\rm H\, I}$ and its uncertainties represented 
by 68\%
confidence intervals by performing a maximum likelihood analysis. 
We find that these uncertainties are underestimated by a factor of 5.75 as compared to the one 
obtained using full simulation grid and covariances.

We repeat this calculation at other redshifts and obtain the $\Gamma_{\rm H\, I}$ and the 68\% 
confidence interval using only the diagonal elements. Next, we multiply this confidence interval 
by factor of 5.75 so that it correctly represents the uncertainties arising from degeneracies in 
the thermal state of the low-$z$ IGM. 
We believe that this approach of obtaining uncertainties at redshifts other than $z=0.2$, although
approximate, is nevertheless a significant improvement over previous estimates of 
$\Gamma_{\rm H\,I}$ that are based on IGM simulation outputs which effectively assume perfect 
knowledge of the thermal state of the IGM. Nevertheless, in a companion paper (Walther et al. in 
prep.), we will present joint constraints on  $T_0$, $\gamma$, and $\Gamma_{\rm H\, I}$. 
For all $\Gamma_{\rm H\, I}$ estimates discussed here, we have fit our power spectrum measurements
over the range $0.03<k<0.1$ s km$^{-1}$, where the smallest $k$ is chosen to minimize the box size
effects and largest $k$ corresponds to scales where we have reliable
uncertainties on the power spectrum measurements.
However, note that our results are not sensitive to the choice of the smallest $k$ value ($k_{\rm 
min}$) as long as $0.03>k_{\rm min}>0.01$ s km$^{-1}$.

\begin{figure}
\includegraphics[width=0.48\textwidth,height=\textheight,keepaspectratio]{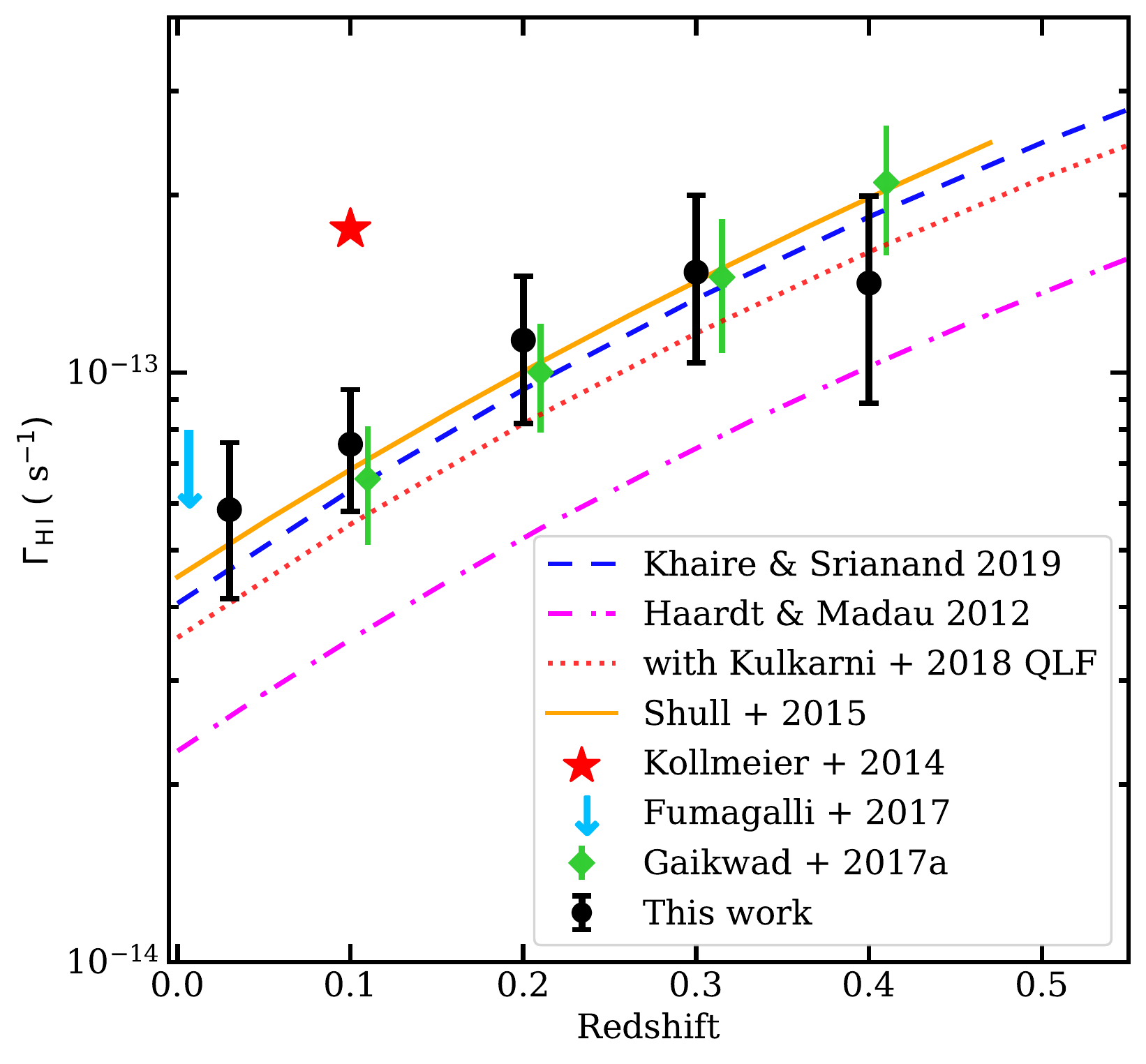}
\caption{Our estimated $\Gamma_{\rm H\, I}$ values (black circles) compared with previous 
measurements by \citet[][green diamonds]{Gaikwad17a}, \citet[][red star]{Kollmeier14}, a fitting 
form by \citet[][orange curve]{Shull15} and upper limits from \citet[][downward 
arrow]{Fumagalli17}. Prediction from a new UV background model of \citet[][blue-dash curve]{KS19} 
is consistent with our measurements. Also, the UV background predictions using updated QLF from 
\citet[][red dotted curve]{Kulkarni18} give consistent results with the measurements (see 
Section~\ref{sec4.2} for more details). The UV background from \citet[][magenta dot-dash 
curve]{HM12} is factor of $\sim 2$ smaller than the measurements.}
\label{fig5}
\end{figure}

\begin{table}
\centering
\caption{$\Gamma_{\rm H\ I}$ measurements}
\begin{tabular}{cccc}
\hline
${\bar z}$& $\Gamma_{\rm H\ I}$ (10$^{-13}$s$^{-1}$) \\
\hline
0.03 &0.585 $^{+0.17}_{-0.18}$\\
0.10 &0.756 $^{+0.17}_{-0.18}$\\
0.20 &1.135 $^{+0.32}_{-0.32}$\\
0.30 &1.479 $^{+0.44}_{-0.52}$\\
0.41 &1.418 $^{+0.53}_{-0.57}$\\ 
 \hline
\end{tabular}
\label{tab3}
\end{table}

Our $\Gamma_{\rm H\, I}$ measurements and the 68\% confidence interval on them
obtained following the method mentioned above are provided in Table~\ref{tab3} and shown 
in Fig.~\ref{fig5}. We also show the previous measurements reported in the 
literature at $z<0.5$ for comparison \citep[however, see][for an early attempt]{Dave01}. The upper
limit at $z\sim 0$ by 
\citet{Fumagalli17} was obtained from the H$\alpha$ and 21-cm observations of a 
nearby galaxy. All other measurements have used the Lyman-$\alpha$ forest from 
\citet{Danforth14}. The measurement by \citet{Kollmeier14} at $z\sim 0.1$ and the 
fit resulting from the analysis by \citet{Shull15} at $z<0.5$ were obtained by 
modeling the CDDF,  whereas measurements of \citet{Gaikwad17a} were obtained using the
flux PDF and power-spectrum (and later confirmed with the CDDF in \citealt{Gaikwad17b}). 
Our measurements are consistent with the previous studies except for 
\citet{Kollmeier14}, which is $\sim 2.5$ times higher. As compared to 
previous studies, we have independent measurements obtained only from the 
Lyman-$\alpha$ forest power spectrum extending down to $z= 0.03$.

In Fig.~\ref{fig5}, we also show predictions from the UV background
models by \citet{HM12} and \citet{KS19}\footnote{For \citet{KS19} UV background, we use their
fiducial Q18 model 
which uses quasar spectral slope $f_{\nu} \propto \nu^{-1.8}$ at hydrogen ionizing 
energies \citep[following][]{Lusso15,Khaire17sed} in the UV background calculations.}. 
Ours as well as the other $\Gamma_{\rm H\, I}$ 
measurements \citep[except for][]{Kollmeier14} are 
consistent with the \citet{KS19} UV background and are a factor of $\sim 2$ higher 
than \citet{HM12}. The \citet{KS19} UV background at $z<0.5$ is mostly 
contributed by quasars since it has been obtained with a negligible contribution 
from galaxies (at $z<2$; see their Eq. 13). Therefore, we argue that quasar emission is sufficient
to produce the low-z UV background \citep[see also]{Khaire15puc}.
The factor of five discrepancy between the measurement from \citet{Kollmeier14} and 
the prediction from \citet{HM12} was argued to represent a photon-underproduction crisis at 
low-$z$. Soon after the \citet{Kollmeier14} study, 
\citet{Khaire15puc} showed that the UV background models that include 
an updated quasar emissivity predict a factor of two higher $\Gamma_{\rm H\, I}$, 
without requiring any additional contribution from galaxies. This model later turned out 
to be consistent with many new $\Gamma_{\rm H\, I}$ measurements 
\citep{Shull15,Gaikwad17a,Gaikwad17b,Gurvich17,Viel17,Fumagalli17}. As shown in 
Fig~\ref{fig5}, this is also consistent with our new measurements. Other recent updates to UV 
background models \citep{Madau15,KS19, Puchwein19} also came to the same 
conclusion using more recent estimates for the  quasar emissivity similar to \citet{Khaire15puc}.

Recently, \citet{Kulkarni18} updated the fits to the quasar luminosity functions (QLF) and 
emissivity across a large redshift range and claim that the quasar contribution to the low-$z$ UV 
background is factor of $\sim 2$
smaller than the predictions by these recent UV background models 
\citep[][]{Madau15, KS19, Puchwein19} and most of the measurements. 
However, the \citet{Kulkarni18} calculation 
uses the CDDF of the IGM from tabulated fits by  \citet{HM12} and an ionizing spectral slope of 
quasars $f_{\nu} \propto \nu^{-1.7}$ \citep{Lusso15}.
Instead, using the \citet{InoueAK14} CDDF results in a 40-60\% higher UV background at $z<0.5$ 
\citep[see, for e.g., Figure 16 of][]{Gaikwad17a}.
This along with including the recombination emissivities in the UV background calculations should 
easily increase the $\Gamma_{\rm H\, I}$ estimates of \citet{Kulkarni18} by factor of $\sim 1.8$ 
as shown in Fig.~\ref{fig5}. 
For this calculation we used the UV background code presented in \citet{KS19} but with the updated
quasar emissivity determined by \citet{Kulkarni18}, with the same limiting magnitude at 1450 
\AA~($M_{\rm 1450, lim}=-18$) and the same spectral slope $f_{\nu} \propto \nu^{-1.7}$ that they
adopted, but using the
\citet{InoueAK14} CDDF instead of the \citet{HM12} fits used by \citet{Kulkarni18}. 
We note that adopting a flatter spectral slope for quasars can further increase the  $\Gamma_{\rm 
H\, I}$. For
example, redoing the calculation mentioned above using the \citet{Kulkarni18} emissivity and the 
\citet{KS19}
UV background model, but changing the quasar spectra slope to $f_{\nu} \propto \nu^{-1.4}$
consistent with the harder slope measured by \citet{Shull12} and \citet{Stevans14}, results in a 
further increase of the $\Gamma_{\rm H\, I}$  by a factor of 1.3.

Finally, note that even if the $\Gamma_{\rm H\, I}$ is as high as that
purported by \citet{Kollmeier14}, there is no crisis associated with
photon production since a negligible contribution from galaxies,
along with updated cosmic star formation histories
\citep{Madau14,Khaire15ebl}, can easily reproduce such a large $\Gamma_{\rm H\, I}$
\citep[see for more details][]{Khaire15puc}.

\section{Summary and conclusions}
We present a new high precision high resolution power spectrum
measurement in five different redshift bins at $z<0.5$. For this measurement, we
have used high-quality medium resolution Lyman-$\alpha$ forest data
from HST/COS from the largest low-$z$ IGM survey published by
\citet{Danforth14}. We applied the procedure developed in
\citet{Walther18a}, which takes into account masked metal-line absorptions, noise in the
data and the finite resolution of the instrument. The data allow us
to reliably probe the power spectrum up to small scales $k<0.1$ s km$^{-1}$
and our measurements show the expected thermal cut-off in the power at
small scales $k>0.03$ s km$^{-1}$, resulting from pressure smoothing of IGM
gas and thermal Doppler broadening of absorption lines. Our power spectrum 
measurements are provided in Table~\ref{tab2}.
 
We compare these with cosmological hydrodynamical simulations 
and obtain constraints of the UV background at $z<0.5$. 
Our measured hydrogen photoionization rates (Table~\ref{tab3}) are consistent with the previous 
estimates \citep{Shull15,Gaikwad17a,Gaikwad17b,Fumagalli17} and recent UV background models 
\citep{KS19, Puchwein19}. This suggests that the low-$z$ UV background is 
dominated by ionizing photons emitted by quasars without requiring any significant 
contribution from galaxies. 

The power-spectrum measurements presented here, in principle, can probe the thermal state of 
the low-$z$ IGM.  At low-$z$, theoretical calculations using standard heating 
and cooling rates show that the IGM loses memory of the previous heating 
episodes caused by the hydrogen and helium reionization. This makes understanding and predicting 
the structure of the low-$z$ IGM relatively simple, as it is independent of the physics
associated with hydrogen and helium reionization heating, which complicate modeling 
at higher redshifts ($z \gtrsim 2$).  In the absence of any other heating processes, theory 
predicts that the diffuse low-density photoionized IGM cools down after $z \sim 2$ 
and asymptotes toward a single temperature-density relation with $\gamma$ close 
to 1.6 and T$_0 \sim 5000$ K \citep{McQuinn16} at $z=0$.
Such a predicted cool-down of 
the IGM at low-$z$ has not yet been observationally confirmed.  
Indeed, there are no reliable measurements of the thermal state of the IGM at $z<1.6$
\citep[but see][]{Ricotti00} where 
the atmospheric cut-off does not allow us to observe Lyman-$\alpha$ forest from 
ground based telescopes.

In a companion paper (Walther et al. in prep.)
we will use the power spectrum measurements presented here
to jointly constrain the IGM thermal state (T$_0$, $\gamma$) and the UV background ($\Gamma_{\rm 
H\, I}$). These measurements will help us understand the physics of IGM and address important 
questions such as whether feedback processes associated with galaxy formation modify the thermal 
state of the IGM at low-$z$ \citep{Viel17, Nasir17} and/or if there is any room for the existence 
of non-standard heating processes powered by TeV Blazars \citep{Puchwein12, Lamberts15} or 
decaying dark matter \citep{Furlanetto06DM_decay,Araya14}.

In contrast with the high-$z$ Universe, the much lower opacity of Lyman series absorption arising 
in the low-$z$ IGM
results in dramatically reduced line blanketing, making it  relatively straightforward to identify
all lines as either resulting from the Lyman-series or metal absorption.
This results in a large redshift path length where Lyman series absorption can be studied, data
analysis and preparation are simplified,  and important systematics from metal-line contamination
arising at small-scales \citep[high-$k$; see section 4.1 of][]{Walther18a} are mitigated.  
Furthermore, our measurements demonstrate that COS resolution is sufficient to
obtain high-quality power-spectrum measurements even at the small-scales (high-$k$, $k\sim 
0.1\,{\rm s\,km^{-1}}$) required for probing the thermal state of the IGM and demonstrate the 
important role that HST/UV spectroscopy can play in our understanding of the low-$z$ IGM. We 
conclude by noting that, owing to the paucity
of archival near-UV spectra covering the Lyman-$\alpha$ transition at $0.5<z<1.6$, there are 
essentially no constraints on the physical state of IGM gas in this redshift interval,
representing 5 Gyr of the Universe's history. It is critical that HST UV spectroscopy fill this 
gap in our understanding of the Universe before HST's mission is complete, otherwise we could 
remain in the dark for decades.

\section*{acknowledgement} 
VK thanks R. Srianand, T. R. Choudhury and P. Gaikwad for insightful discussion on the power 
spectrum normalization. We thank all members of the ENIGMA 
group\footnote{\url{http://enigma.physics.ucsb.edu/}} at University of California Santa Barbara 
for useful discussions and suggestions.

Financial support for this work was provided to VK by NASA
through grant number HST-AR-15032.002-A from the Space Telescope Science Institute, which is 
operated by Associated Universities for Research in Astronomy, Inc., under NASA contract NAS
5-26555.

Calculations presented in this paper used the draco  and hydra clusters of the Max Planck 
Computing and Data Facility, a center of the Max Planck Society in Garching (Germany).
We have also used the National Energy Research Scientific Computing Center (NERSC) supported by 
the U.S. Department of Energy (DoE) under Contract No. DE-AC02-05CH11231. ZL was in part supported
by the Scientific Discovery through Advanced Computing (SciDAC) program funded by the DoE, the 
Office of High Energy Physics and the  Office of Advanced Scientific Computing Research.

\bibliographystyle{mnras}
\bibliography{vikrambib}

\begin{thebibliography}{}
\makeatletter
\relax
\def\mn@urlcharsother{\let\do\@makeother \do\$\do\&\do\#\do\^\do\_\do\%\do\~}
\def\mn@doi{\begingroup\mn@urlcharsother \@ifnextchar [ {\mn@doi@}
  {\mn@doi@[]}}
\def\mn@doi@[#1]#2{\def\@tempa{#1}\ifx\@tempa\@empty \href
  {http://dx.doi.org/#2} {doi:#2}\else \href {http://dx.doi.org/#2} {#1}\fi
  \endgroup}
\def\mn@eprint#1#2{\mn@eprint@#1:#2::\@nil}
\def\mn@eprint@arXiv#1{\href {http://arxiv.org/abs/#1} {{\tt arXiv:#1}}}
\def\mn@eprint@dblp#1{\href {http://dblp.uni-trier.de/rec/bibtex/#1.xml}
  {dblp:#1}}
\def\mn@eprint@#1:#2:#3:#4\@nil{\def\@tempa {#1}\def\@tempb {#2}\def\@tempc
  {#3}\ifx \@tempc \@empty \let \@tempc \@tempb \let \@tempb \@tempa \fi \ifx
  \@tempb \@empty \def\@tempb {arXiv}\fi \@ifundefined
  {mn@eprint@\@tempb}{\@tempb:\@tempc}{\expandafter \expandafter \csname
  mn@eprint@\@tempb\endcsname \expandafter{\@tempc}}}

\bibitem[\protect\citeauthoryear{{Almgren}, {Bell}, {Lijewski}, {Luki{\'c}}  \&
  {Van Andel}}{{Almgren} et~al.}{2013}]{Almgren13}
{Almgren} A.~S.,  {Bell} J.~B.,  {Lijewski} M.~J.,  {Luki{\'c}} Z.,   {Van
  Andel} E.,  2013, \mn@doi [\apj] {10.1088/0004-637X/765/1/39}, \href
  {http://adsabs.harvard.edu/abs/2013ApJ...765...39A} {765, 39}

\bibitem[\protect\citeauthoryear{{Araya} \& {Padilla}}{{Araya} \&
  {Padilla}}{2014}]{Araya14}
{Araya} I.~J.,  {Padilla} N.~D.,  2014, \mn@doi [\mnras]
  {10.1093/mnras/stu1780}, \href
  {http://adsabs.harvard.edu/abs/2014MNRAS.445..850A} {445, 850}

\bibitem[\protect\citeauthoryear{{Becker}, {Bolton}, {Haehnelt}  \&
  {Sargent}}{{Becker} et~al.}{2011}]{Becker11t}
{Becker} G.~D.,  {Bolton} J.~S.,  {Haehnelt} M.~G.,   {Sargent} W.~L.~W.,
  2011, \mn@doi [\mnras] {10.1111/j.1365-2966.2010.17507.x}, \href
  {http://adsabs.harvard.edu/abs/2011MNRAS.410.1096B} {410, 1096}

\bibitem[\protect\citeauthoryear{{Bolton}, {Becker}, {Raskutti}, {Wyithe},
  {Haehnelt}  \& {Sargent}}{{Bolton} et~al.}{2012}]{Bolton12}
{Bolton} J.~S.,  {Becker} G.~D.,  {Raskutti} S.,  {Wyithe} J.~S.~B.,
  {Haehnelt} M.~G.,   {Sargent} W.~L.~W.,  2012, \mn@doi [\mnras]
  {10.1111/j.1365-2966.2011.19929.x}, \href
  {http://adsabs.harvard.edu/abs/2012MNRAS.419.2880B} {419, 2880}

\bibitem[\protect\citeauthoryear{{Borthakur} et~al.,}{{Borthakur}
  et~al.}{2015}]{Borthakur15}
{Borthakur} S.,  et~al., 2015, \mn@doi [\apj] {10.1088/0004-637X/813/1/46},
  \href {http://adsabs.harvard.edu/abs/2015ApJ...813...46B} {813, 46}

\bibitem[\protect\citeauthoryear{{Burchett} et~al.,}{{Burchett}
  et~al.}{2015}]{Burchett15}
{Burchett} J.~N.,  et~al., 2015, \mn@doi [\apj] {10.1088/0004-637X/815/2/91},
  \href {http://adsabs.harvard.edu/abs/2015ApJ...815...91B} {815, 91}

\bibitem[\protect\citeauthoryear{{Croft}, {Weinberg}, {Katz}  \&
  {Hernquist}}{{Croft} et~al.}{1998}]{Croft98}
{Croft} R.~A.~C.,  {Weinberg} D.~H.,  {Katz} N.,   {Hernquist} L.,  1998,
  \mn@doi [\apj] {10.1086/305289}, \href
  {http://adsabs.harvard.edu/abs/1998ApJ...495...44C} {495, 44}

\bibitem[\protect\citeauthoryear{{Croft}, {Weinberg}, {Bolte}, {Burles},
  {Hernquist}, {Katz}, {Kirkman}  \& {Tytler}}{{Croft} et~al.}{2002}]{Croft02}
{Croft} R.~A.~C.,  {Weinberg} D.~H.,  {Bolte} M.,  {Burles} S.,  {Hernquist}
  L.,  {Katz} N.,  {Kirkman} D.,   {Tytler} D.,  2002, \mn@doi [\apj]
  {10.1086/344099}, \href {http://adsabs.harvard.edu/abs/2002ApJ...581...20C}
  {581, 20}

\bibitem[\protect\citeauthoryear{{Danforth} et~al.,}{{Danforth}
  et~al.}{2016}]{Danforth14}
{Danforth} C.~W.,  et~al., 2016, \mn@doi [\apj] {10.3847/0004-637X/817/2/111},
  \href {http://adsabs.harvard.edu/abs/2016ApJ...817..111D} {817, 111}

\bibitem[\protect\citeauthoryear{{Dav{\'e}} \& {Tripp}}{{Dav{\'e}} \&
  {Tripp}}{2001}]{Dave01}
{Dav{\'e}} R.,  {Tripp} T.~M.,  2001, \mn@doi [\apj] {10.1086/320977}, \href
  {http://adsabs.harvard.edu/abs/2001ApJ...553..528D} {553, 528}

\bibitem[\protect\citeauthoryear{{Dav{\'e}}, {Oppenheimer}, {Katz}, {Kollmeier}
   \& {Weinberg}}{{Dav{\'e}} et~al.}{2010}]{Dave10}
{Dav{\'e}} R.,  {Oppenheimer} B.~D.,  {Katz} N.,  {Kollmeier} J.~A.,
  {Weinberg} D.~H.,  2010, \mn@doi [\mnras] {10.1111/j.1365-2966.2010.17279.x},
  \href {http://cdsads.u-strasbg.fr/abs/2010MNRAS.408.2051D} {408, 2051}

\bibitem[\protect\citeauthoryear{{Fumagalli}, {Haardt}, {Theuns}, {Morris},
  {Cantalupo}, {Madau}  \& {Fossati}}{{Fumagalli} et~al.}{2017}]{Fumagalli17}
{Fumagalli} M.,  {Haardt} F.,  {Theuns} T.,  {Morris} S.~L.,  {Cantalupo} S.,
  {Madau} P.,   {Fossati} M.,  2017, \mn@doi [\mnras] {10.1093/mnras/stx398},
  \href {http://adsabs.harvard.edu/abs/2017MNRAS.467.4802F} {467, 4802}

\bibitem[\protect\citeauthoryear{{Furlanetto}, {Oh}  \&
  {Pierpaoli}}{{Furlanetto} et~al.}{2006}]{Furlanetto06DM_decay}
{Furlanetto} S.~R.,  {Oh} S.~P.,   {Pierpaoli} E.,  2006, \mn@doi [\prd]
  {10.1103/PhysRevD.74.103502}, \href
  {http://adsabs.harvard.edu/abs/2006PhRvD..74j3502F} {74, 103502}

\bibitem[\protect\citeauthoryear{{Gaikwad}, {Khaire}, {Choudhury}  \&
  {Srianand}}{{Gaikwad} et~al.}{2017a}]{Gaikwad17a}
{Gaikwad} P.,  {Khaire} V.,  {Choudhury} T.~R.,   {Srianand} R.,  2017a,
  \mn@doi [\mnras] {10.1093/mnras/stw3086}, \href
  {http://adsabs.harvard.edu/abs/2017MNRAS.466..838G} {466, 838}

\bibitem[\protect\citeauthoryear{{Gaikwad}, {Srianand}, {Choudhury}  \&
  {Khaire}}{{Gaikwad} et~al.}{2017b}]{Gaikwad17b}
{Gaikwad} P.,  {Srianand} R.,  {Choudhury} T.~R.,   {Khaire} V.,  2017b,
  \mn@doi [\mnras] {10.1093/mnras/stx248}, \href
  {http://adsabs.harvard.edu/abs/2017MNRAS.467.3172G} {467, 3172}

\bibitem[\protect\citeauthoryear{{Garzilli}, {Boyarsky}  \&
  {Ruchayskiy}}{{Garzilli} et~al.}{2017}]{Garzilli17}
{Garzilli} A.,  {Boyarsky} A.,   {Ruchayskiy} O.,  2017, \mn@doi [Physics
  Letters B] {10.1016/j.physletb.2017.08.022}, \href
  {http://adsabs.harvard.edu/abs/2017PhLB..773..258G} {773, 258}

\bibitem[\protect\citeauthoryear{{Gunn} \& {Peterson}}{{Gunn} \&
  {Peterson}}{1965}]{Gunn65}
{Gunn} J.~E.,  {Peterson} B.~A.,  1965, \mn@doi [\apj] {10.1086/148444}, \href
  {http://adsabs.harvard.edu/abs/1965ApJ...142.1633G} {142, 1633}

\bibitem[\protect\citeauthoryear{{Gurvich}, {Burkhart}  \& {Bird}}{{Gurvich}
  et~al.}{2017}]{Gurvich17}
{Gurvich} A.,  {Burkhart} B.,   {Bird} S.,  2017, \mn@doi [\apj]
  {10.3847/1538-4357/835/2/175}, \href
  {http://adsabs.harvard.edu/abs/2017ApJ...835..175G} {835, 175}

\bibitem[\protect\citeauthoryear{{Haardt} \& {Madau}}{{Haardt} \&
  {Madau}}{2012}]{HM12}
{Haardt} F.,  {Madau} P.,  2012, \mn@doi [\apj] {10.1088/0004-637X/746/2/125},
  \href {http://adsabs.harvard.edu/abs/2012ApJ...746..125H} {746, 125}

\bibitem[\protect\citeauthoryear{{Haehnelt} \& {Steinmetz}}{{Haehnelt} \&
  {Steinmetz}}{1998}]{Haehnelt98}
{Haehnelt} M.~G.,  {Steinmetz} M.,  1998, \mn@doi [\mnras]
  {10.1046/j.1365-8711.1998.01879.x}, \href
  {http://adsabs.harvard.edu/abs/1998MNRAS.298L..21H} {298, L21}

\bibitem[\protect\citeauthoryear{{Hahn} \& {Abel}}{{Hahn} \&
  {Abel}}{2011}]{Hahn11}
{Hahn} O.,  {Abel} T.,  2011, \mn@doi [\mnras]
  {10.1111/j.1365-2966.2011.18820.x}, \href
  {http://adsabs.harvard.edu/abs/2011MNRAS.415.2101H} {415, 2101}

\bibitem[\protect\citeauthoryear{{Hiss}, {Walther}, {Hennawi}, {O{\~n}orbe},
  {O'Meara}, {Rorai}  \& {Luki{\'c}}}{{Hiss} et~al.}{2018}]{Hiss18}
{Hiss} H.,  {Walther} M.,  {Hennawi} J.~F.,  {O{\~n}orbe} J.,  {O'Meara} J.~M.,
   {Rorai} A.,   {Luki{\'c}} Z.,  2018, \mn@doi [\apj]
  {10.3847/1538-4357/aada86}, \href
  {http://adsabs.harvard.edu/abs/2018ApJ...865...42H} {865, 42}

\bibitem[\protect\citeauthoryear{{Howlett}, {Lewis}, {Hall}  \&
  {Challinor}}{{Howlett} et~al.}{2012}]{Howlett12}
{Howlett} C.,  {Lewis} A.,  {Hall} A.,   {Challinor} A.,  2012, \mn@doi [\jcap]
  {10.1088/1475-7516/2012/04/027}, \href
  {http://adsabs.harvard.edu/abs/2012JCAP...04..027H} {4, 027}

\bibitem[\protect\citeauthoryear{{Hui} \& {Gnedin}}{{Hui} \&
  {Gnedin}}{1997}]{Hui97}
{Hui} L.,  {Gnedin} N.~Y.,  1997, \mn@doi [\mnras] {10.1093/mnras/292.1.27},
  \href {http://adsabs.harvard.edu/abs/1997MNRAS.292...27H} {292, 27}

\bibitem[\protect\citeauthoryear{{Inoue}, {Shimizu}, {Iwata}  \&
  {Tanaka}}{{Inoue} et~al.}{2014}]{InoueAK14}
{Inoue} A.~K.,  {Shimizu} I.,  {Iwata} I.,   {Tanaka} M.,  2014, \mn@doi
  [\mnras] {10.1093/mnras/stu936}, \href
  {http://adsabs.harvard.edu/abs/2014MNRAS.442.1805I} {442, 1805}

\bibitem[\protect\citeauthoryear{{Irsic} et~al.,}{{Irsic}
  et~al.}{2017a}]{Irsic17t}
{Irsic} V.,  et~al., 2017a, \mn@doi [Physical Review D]
  {10.1103/PhysRevD.96.023522}, \href
  {http://adsabs.harvard.edu/abs/2017PhRvD..96b3522I} {96, 023522}

\bibitem[\protect\citeauthoryear{{Irsic}, {Viel}, {Haehnelt}, {Bolton}  \&
  {Becker}}{{Irsic} et~al.}{2017b}]{Irsic17}
{Irsic} V.,  {Viel} M.,  {Haehnelt} M.~G.,  {Bolton} J.~S.,   {Becker} G.~D.,
  2017b, \mn@doi [Physical Review Letters] {10.1103/PhysRevLett.119.031302},
  \href {http://adsabs.harvard.edu/abs/2017PhRvL.119c1302I} {119, 031302}

\bibitem[\protect\citeauthoryear{{Khaire}}{{Khaire}}{2017}]{Khaire17sed}
{Khaire} V.,  2017, \mn@doi [\mnras] {10.1093/mnras/stx1487}, \href
  {http://adsabs.harvard.edu/abs/2017MNRAS.471..255K} {471, 255}

\bibitem[\protect\citeauthoryear{{Khaire} \& {Srianand}}{{Khaire} \&
  {Srianand}}{2015a}]{Khaire15puc}
{Khaire} V.,  {Srianand} R.,  2015a, \mn@doi [\mnras] {10.1093/mnrasl/slv060},
  \href {http://adsabs.harvard.edu/abs/2015MNRAS.451L..30K} {451, L30}

\bibitem[\protect\citeauthoryear{{Khaire} \& {Srianand}}{{Khaire} \&
  {Srianand}}{2015b}]{Khaire15ebl}
{Khaire} V.,  {Srianand} R.,  2015b, \mn@doi [\apj]
  {10.1088/0004-637X/805/1/33}, \href
  {http://adsabs.harvard.edu/abs/2015ApJ...805...33K} {805, 33}

\bibitem[\protect\citeauthoryear{{Khaire} \& {Srianand}}{{Khaire} \&
  {Srianand}}{2019}]{KS19}
{Khaire} V.,  {Srianand} R.,  2019, \mn@doi [\mnras] {10.1093/mnras/stz174},
  \href {http://adsabs.harvard.edu/abs/2019MNRAS.484.4174K} {484, 4174}

\bibitem[\protect\citeauthoryear{{Kim}, {Viel}, {Haehnelt}, {Carswell}  \&
  {Cristiani}}{{Kim} et~al.}{2004}]{Kim04ps}
{Kim} T.-S.,  {Viel} M.,  {Haehnelt} M.~G.,  {Carswell} R.~F.,   {Cristiani}
  S.,  2004, \mn@doi [\mnras] {10.1111/j.1365-2966.2004.07221.x}, \href
  {http://adsabs.harvard.edu/abs/2004MNRAS.347..355K} {347, 355}

\bibitem[\protect\citeauthoryear{{Kollmeier} et~al.,}{{Kollmeier}
  et~al.}{2014}]{Kollmeier14}
{Kollmeier} J.~A.,  et~al., 2014, \mn@doi [\apjl]
  {10.1088/2041-8205/789/2/L32}, \href
  {http://adsabs.harvard.edu/abs/2014ApJ...789L..32K} {789, L32}

\bibitem[\protect\citeauthoryear{{Kulkarni}, {Hennawi}, {O{\~n}orbe}, {Rorai}
  \& {Springel}}{{Kulkarni} et~al.}{2015}]{Kulkarni15}
{Kulkarni} G.,  {Hennawi} J.~F.,  {O{\~n}orbe} J.,  {Rorai} A.,   {Springel}
  V.,  2015, \mn@doi [\apj] {10.1088/0004-637X/812/1/30}, \href
  {http://adsabs.harvard.edu/abs/2015ApJ...812...30K} {812, 30}

\bibitem[\protect\citeauthoryear{{Kulkarni}, {Worseck}  \&
  {Hennawi}}{{Kulkarni} et~al.}{2018}]{Kulkarni18}
{Kulkarni} G.,  {Worseck} G.,   {Hennawi} J.~F.,  2018, preprint, \href
  {http://adsabs.harvard.edu/abs/2018arXiv180709774K} {} (\mn@eprint {arXiv}
  {1807.09774})

\bibitem[\protect\citeauthoryear{{Lamberts}, {Chang}, {Pfrommer}, {Puchwein},
  {Broderick}  \& {Shalaby}}{{Lamberts} et~al.}{2015}]{Lamberts15}
{Lamberts} A.,  {Chang} P.,  {Pfrommer} C.,  {Puchwein} E.,  {Broderick} A.~E.,
    {Shalaby} M.,  2015, \mn@doi [\apj] {10.1088/0004-637X/811/1/19}, \href
  {http://adsabs.harvard.edu/abs/2015ApJ...811...19L} {811, 19}

\bibitem[\protect\citeauthoryear{{Lewis}, {Challinor}  \& {Lasenby}}{{Lewis}
  et~al.}{2000}]{Lewis00}
{Lewis} A.,  {Challinor} A.,   {Lasenby} A.,  2000, \mn@doi [\apj]
  {10.1086/309179}, \href {http://adsabs.harvard.edu/abs/2000ApJ...538..473L}
  {538, 473}

\bibitem[\protect\citeauthoryear{{Lidz}, {Faucher-Gigu{\`e}re}, {Dall'Aglio},
  {McQuinn}, {Fechner}, {Zaldarriaga}, {Hernquist}  \& {Dutta}}{{Lidz}
  et~al.}{2010}]{Lidz10}
{Lidz} A.,  {Faucher-Gigu{\`e}re} C.-A.,  {Dall'Aglio} A.,  {McQuinn} M.,
  {Fechner} C.,  {Zaldarriaga} M.,  {Hernquist} L.,   {Dutta} S.,  2010,
  \mn@doi [\apj] {10.1088/0004-637X/718/1/199}, \href
  {http://adsabs.harvard.edu/abs/2010ApJ...718..199L} {718, 199}

\bibitem[\protect\citeauthoryear{{Lomb}}{{Lomb}}{1976}]{Lomb76}
{Lomb} N.~R.,  1976, \mn@doi [\apss] {10.1007/BF00648343}, \href
  {http://adsabs.harvard.edu/abs/1976Ap%26SS..39..447L} {39, 447}

\bibitem[\protect\citeauthoryear{{Luki{\'c}}, {Stark}, {Nugent}, {White},
  {Meiksin}  \& {Almgren}}{{Luki{\'c}} et~al.}{2015}]{Lukic15}
{Luki{\'c}} Z.,  {Stark} C.~W.,  {Nugent} P.,  {White} M.,  {Meiksin} A.~A.,
  {Almgren} A.,  2015, \mn@doi [\mnras] {10.1093/mnras/stu2377}, \href
  {http://adsabs.harvard.edu/abs/2015MNRAS.446.3697L} {446, 3697}

\bibitem[\protect\citeauthoryear{{Lusso}, {Worseck}, {Hennawi}, {Prochaska},
  {Vignali}, {Stern}  \& {O'Meara}}{{Lusso} et~al.}{2015}]{Lusso15}
{Lusso} E.,  {Worseck} G.,  {Hennawi} J.~F.,  {Prochaska} J.~X.,  {Vignali} C.,
   {Stern} J.,   {O'Meara} J.~M.,  2015, \mn@doi [\mnras]
  {10.1093/mnras/stv516}, \href
  {http://adsabs.harvard.edu/abs/2015MNRAS.449.4204L} {449, 4204}

\bibitem[\protect\citeauthoryear{{Madau} \& {Dickinson}}{{Madau} \&
  {Dickinson}}{2014}]{Madau14}
{Madau} P.,  {Dickinson} M.,  2014, \mn@doi [\araa]
  {10.1146/annurev-astro-081811-125615}, \href
  {http://adsabs.harvard.edu/abs/2014ARA%26A..52..415M} {52, 415}

\bibitem[\protect\citeauthoryear{{Madau} \& {Haardt}}{{Madau} \&
  {Haardt}}{2015}]{Madau15}
{Madau} P.,  {Haardt} F.,  2015, \mn@doi [\apjl] {10.1088/2041-8205/813/1/L8},
  \href {http://adsabs.harvard.edu/abs/2015ApJ...813L...8M} {813, L8}

\bibitem[\protect\citeauthoryear{{McDonald}, {Miralda-Escud{\'e}}, {Rauch},
  {Sargent}, {Barlow}, {Cen}  \& {Ostriker}}{{McDonald}
  et~al.}{2000}]{McDonald00}
{McDonald} P.,  {Miralda-Escud{\'e}} J.,  {Rauch} M.,  {Sargent} W.~L.~W.,
  {Barlow} T.~A.,  {Cen} R.,   {Ostriker} J.~P.,  2000, \mn@doi [\apj]
  {10.1086/317079}, \href {http://adsabs.harvard.edu/abs/2000ApJ...543....1M}
  {543, 1}

\bibitem[\protect\citeauthoryear{{McDonald} et~al.,}{{McDonald}
  et~al.}{2006}]{McDonald06}
{McDonald} P.,  et~al., 2006, \mn@doi [\apjs] {10.1086/444361}, \href
  {http://adsabs.harvard.edu/abs/2006ApJS..163...80M} {163, 80}

\bibitem[\protect\citeauthoryear{{McQuinn}}{{McQuinn}}{2016}]{McQuinn16}
{McQuinn} M.,  2016, \mn@doi [\araa] {10.1146/annurev-astro-082214-122355},
  \href {http://adsabs.harvard.edu/abs/2016ARA%26A..54..313M} {54, 313}

\bibitem[\protect\citeauthoryear{{Nasir}, {Bolton}, {Viel}, {Kim}, {Haehnelt},
  {Puchwein}  \& {Sijacki}}{{Nasir} et~al.}{2017}]{Nasir17}
{Nasir} F.,  {Bolton} J.~S.,  {Viel} M.,  {Kim} T.-S.,  {Haehnelt} M.~G.,
  {Puchwein} E.,   {Sijacki} D.,  2017, \mn@doi [\mnras]
  {10.1093/mnras/stx1648}, \href
  {http://cdsads.u-strasbg.fr/abs/2017MNRAS.471.1056N} {471, 1056}

\bibitem[\protect\citeauthoryear{{O{\~n}orbe}, {Hennawi}  \&
  {Luki{\'c}}}{{O{\~n}orbe} et~al.}{2017}]{Onorbe17}
{O{\~n}orbe} J.,  {Hennawi} J.~F.,   {Luki{\'c}} Z.,  2017, \mn@doi [\apj]
  {10.3847/1538-4357/aa6031}, \href
  {http://adsabs.harvard.edu/abs/2017ApJ...837..106O} {837, 106}

\bibitem[\protect\citeauthoryear{{Palanque-Delabrouille}
  et~al.,}{{Palanque-Delabrouille} et~al.}{2013}]{Palanque13ps}
{Palanque-Delabrouille} N.,  et~al., 2013, \mn@doi [\aap]
  {10.1051/0004-6361/201322130}, \href
  {http://adsabs.harvard.edu/abs/2013A%26A...559A..85P} {559, A85}

\bibitem[\protect\citeauthoryear{{Palanque-Delabrouille}
  et~al.,}{{Palanque-Delabrouille} et~al.}{2015}]{PD15Neutrino}
{Palanque-Delabrouille} N.,  et~al., 2015, \mn@doi [\jcap]
  {10.1088/1475-7516/2015/11/011}, \href
  {http://adsabs.harvard.edu/abs/2015JCAP...11..011P} {11, 011}

\bibitem[\protect\citeauthoryear{{Peeples}, {Weinberg}, {Dav{\'e}}, {Fardal}
  \& {Katz}}{{Peeples} et~al.}{2010}]{Peeples10}
{Peeples} M.~S.,  {Weinberg} D.~H.,  {Dav{\'e}} R.,  {Fardal} M.~A.,   {Katz}
  N.,  2010, \mn@doi [\mnras] {10.1111/j.1365-2966.2010.16383.x}, \href
  {http://adsabs.harvard.edu/abs/2010MNRAS.404.1281P} {404, 1281}

\bibitem[\protect\citeauthoryear{{Planck Collaboration} et~al.,}{{Planck
  Collaboration} et~al.}{2018}]{Planck18}
{Planck Collaboration} et~al., 2018, preprint, \href
  {http://adsabs.harvard.edu/abs/2018arXiv180706209P} {} (\mn@eprint {arXiv}
  {1807.06209})

\bibitem[\protect\citeauthoryear{{Puchwein}, {Pfrommer}, {Springel},
  {Broderick}  \& {Chang}}{{Puchwein} et~al.}{2012}]{Puchwein12}
{Puchwein} E.,  {Pfrommer} C.,  {Springel} V.,  {Broderick} A.~E.,   {Chang}
  P.,  2012, \mn@doi [\mnras] {10.1111/j.1365-2966.2012.20738.x}, \href
  {http://adsabs.harvard.edu/abs/2012MNRAS.423..149P} {423, 149}

\bibitem[\protect\citeauthoryear{{Puchwein}, {Haardt}, {Haehnelt}  \&
  {Madau}}{{Puchwein} et~al.}{2019}]{Puchwein19}
{Puchwein} E.,  {Haardt} F.,  {Haehnelt} M.~G.,   {Madau} P.,  2019, \mn@doi
  [\mnras] {10.1093/mnras/stz222}, \href
  {http://adsabs.harvard.edu/abs/2019MNRAS.485...47P} {485, 47}

\bibitem[\protect\citeauthoryear{{Rahmati}, {Pawlik}, {Raicevic}  \&
  {Schaye}}{{Rahmati} et~al.}{2013}]{Rahmati13}
{Rahmati} A.,  {Pawlik} A.~H.,  {Raicevic} M.,   {Schaye} J.,  2013, \mn@doi
  [\mnras] {10.1093/mnras/stt066}, \href
  {http://adsabs.harvard.edu/abs/2013MNRAS.430.2427R} {430, 2427}

\bibitem[\protect\citeauthoryear{{Ricotti} \& {Shull}}{{Ricotti} \&
  {Shull}}{2000}]{Ricotti00}
{Ricotti} M.,  {Shull} J.~M.,  2000, \mn@doi [\apj] {10.1086/317025}, \href
  {http://adsabs.harvard.edu/abs/2000ApJ...542..548R} {542, 548}

\bibitem[\protect\citeauthoryear{{Rorai}, {Hennawi}  \& {White}}{{Rorai}
  et~al.}{2013}]{Rorai13}
{Rorai} A.,  {Hennawi} J.~F.,   {White} M.,  2013, \mn@doi [\apj]
  {10.1088/0004-637X/775/2/81}, \href
  {http://adsabs.harvard.edu/abs/2013ApJ...775...81R} {775, 81}

\bibitem[\protect\citeauthoryear{{Rorai} et~al.,}{{Rorai}
  et~al.}{2017}]{Rorai17}
{Rorai} A.,  et~al., 2017, \mn@doi [Science] {10.1126/science.aaf9346}, \href
  {http://adsabs.harvard.edu/abs/2017Sci...356..418R} {356, 418}

\bibitem[\protect\citeauthoryear{{Scargle}}{{Scargle}}{1982}]{Scargle82}
{Scargle} J.~D.,  1982, \mn@doi [\apj] {10.1086/160554}, \href
  {http://adsabs.harvard.edu/abs/1982ApJ...263..835S} {263, 835}

\bibitem[\protect\citeauthoryear{{Schaye}, {Theuns}, {Leonard}  \&
  {Efstathiou}}{{Schaye} et~al.}{1999}]{Schaye99}
{Schaye} J.,  {Theuns} T.,  {Leonard} A.,   {Efstathiou} G.,  1999, \mn@doi
  [\mnras] {10.1046/j.1365-8711.1999.02956.x}, \href
  {http://adsabs.harvard.edu/abs/1999MNRAS.310...57S} {310, 57}

\bibitem[\protect\citeauthoryear{{Shull}, {Harness}, {Trenti}  \&
  {Smith}}{{Shull} et~al.}{2012}]{Shull12}
{Shull} J.~M.,  {Harness} A.,  {Trenti} M.,   {Smith} B.~D.,  2012, \mn@doi
  [\apj] {10.1088/0004-637X/747/2/100}, \href
  {http://adsabs.harvard.edu/abs/2012ApJ...747..100S} {747, 100}

\bibitem[\protect\citeauthoryear{{Shull}, {Moloney}, {Danforth}  \&
  {Tilton}}{{Shull} et~al.}{2015}]{Shull15}
{Shull} J.~M.,  {Moloney} J.,  {Danforth} C.~W.,   {Tilton} E.~M.,  2015,
  \mn@doi [\apj] {10.1088/0004-637X/811/1/3}, \href
  {http://adsabs.harvard.edu/abs/2015ApJ...811....3S} {811, 3}

\bibitem[\protect\citeauthoryear{{Stevans}, {Shull}, {Danforth}  \&
  {Tilton}}{{Stevans} et~al.}{2014}]{Stevans14}
{Stevans} M.~L.,  {Shull} J.~M.,  {Danforth} C.~W.,   {Tilton} E.~M.,  2014,
  \mn@doi [\apj] {10.1088/0004-637X/794/1/75}, \href
  {http://adsabs.harvard.edu/abs/2014ApJ...794...75S} {794, 75}

\bibitem[\protect\citeauthoryear{{Tepper-Garc{\'{\i}}a}}{{Tepper-Garc{\'{\i}}a}}{2006}]{Tepper06}
{Tepper-Garc{\'{\i}}a} T.,  2006, \mn@doi [\mnras]
  {10.1111/j.1365-2966.2006.10450.x}, \href
  {http://adsabs.harvard.edu/abs/2006MNRAS.369.2025T} {369, 2025}

\bibitem[\protect\citeauthoryear{{Theuns}, {Leonard}, {Efstathiou}, {Pearce}
  \& {Thomas}}{{Theuns} et~al.}{1998}]{Theuns98}
{Theuns} T.,  {Leonard} A.,  {Efstathiou} G.,  {Pearce} F.~R.,   {Thomas}
  P.~A.,  1998, \mn@doi [\mnras] {10.1046/j.1365-8711.1998.02040.x}, \href
  {http://adsabs.harvard.edu/abs/1998MNRAS.301..478T} {301, 478}

\bibitem[\protect\citeauthoryear{{Theuns}, {Schaye}  \& {Haehnelt}}{{Theuns}
  et~al.}{2000}]{Theuns00}
{Theuns} T.,  {Schaye} J.,   {Haehnelt} M.~G.,  2000, \mn@doi [\mnras]
  {10.1046/j.1365-8711.2000.03423.x}, \href
  {http://adsabs.harvard.edu/abs/2000MNRAS.315..600T} {315, 600}

\bibitem[\protect\citeauthoryear{{Tumlinson} et~al.,}{{Tumlinson}
  et~al.}{2013}]{Tumlinson13}
{Tumlinson} J.,  et~al., 2013, \mn@doi [\apj] {10.1088/0004-637X/777/1/59},
  \href {http://adsabs.harvard.edu/abs/2013ApJ...777...59T} {777, 59}

\bibitem[\protect\citeauthoryear{{Upton Sanderbeck}, {D'Aloisio}  \&
  {McQuinn}}{{Upton Sanderbeck} et~al.}{2016}]{Pheobe16t0}
{Upton Sanderbeck} P.~R.,  {D'Aloisio} A.,   {McQuinn} M.~J.,  2016, \mn@doi
  [\mnras] {10.1093/mnras/stw1117}, \href
  {http://adsabs.harvard.edu/abs/2016MNRAS.460.1885U} {460, 1885}

\bibitem[\protect\citeauthoryear{{Viel}, {Becker}, {Bolton}, {Haehnelt},
  {Rauch}  \& {Sargent}}{{Viel} et~al.}{2008}]{Viel08}
{Viel} M.,  {Becker} G.~D.,  {Bolton} J.~S.,  {Haehnelt} M.~G.,  {Rauch} M.,
  {Sargent} W.~L.~W.,  2008, \mn@doi [Physical Review Letters]
  {10.1103/PhysRevLett.100.041304}, \href
  {http://adsabs.harvard.edu/abs/2008PhRvL.100d1304V} {100, 041304}

\bibitem[\protect\citeauthoryear{{Viel}, {Becker}, {Bolton}  \&
  {Haehnelt}}{{Viel} et~al.}{2013}]{Viel13}
{Viel} M.,  {Becker} G.~D.,  {Bolton} J.~S.,   {Haehnelt} M.~G.,  2013, \mn@doi
  [\prd] {10.1103/PhysRevD.88.043502}, \href
  {http://adsabs.harvard.edu/abs/2013PhRvD..88d3502V} {88, 043502}

\bibitem[\protect\citeauthoryear{{Viel}, {Haehnelt}, {Bolton}, {Kim},
  {Puchwein}, {Nasir}  \& {Wakker}}{{Viel} et~al.}{2017}]{Viel17}
{Viel} M.,  {Haehnelt} M.~G.,  {Bolton} J.~S.,  {Kim} T.-S.,  {Puchwein} E.,
  {Nasir} F.,   {Wakker} B.~P.,  2017, \mn@doi [\mnras]
  {10.1093/mnrasl/slx004}, \href
  {http://adsabs.harvard.edu/abs/2017MNRAS.467L..86V} {467, L86}

\bibitem[\protect\citeauthoryear{{Wakker}, {Hernandez}, {French}, {Kim},
  {Oppenheimer}  \& {Savage}}{{Wakker} et~al.}{2015}]{Wakker15}
{Wakker} B.~P.,  {Hernandez} A.~K.,  {French} D.~M.,  {Kim} T.-S.,
  {Oppenheimer} B.~D.,   {Savage} B.~D.,  2015, \mn@doi [\apj]
  {10.1088/0004-637X/814/1/40}, \href
  {http://adsabs.harvard.edu/abs/2015ApJ...814...40W} {814, 40}

\bibitem[\protect\citeauthoryear{{Walther}, {Hennawi}, {Hiss}, {O{\~n}orbe},
  {Lee}  \& {Rorai}}{{Walther} et~al.}{2018}]{Walther18a}
{Walther} M.,  {Hennawi} J.~F.,  {Hiss} H.,  {O{\~n}orbe} J.,  {Lee} K.-G.,
  {Rorai} A.,  2018, \mn@doi [\apj] {10.3847/1538-4357/aa9c81}, \href
  {http://adsabs.harvard.edu/abs/2018ApJ...852...22W} {852, 22}

\bibitem[\protect\citeauthoryear{{Walther}, {O{\~n}orbe}, {Hennawi}  \&
  {Luki{\'c}}}{{Walther} et~al.}{2019}]{Walther19}
{Walther} M.,  {O{\~n}orbe} J.,  {Hennawi} J.~F.,   {Luki{\'c}} Z.,  2019,
  \mn@doi [\apj] {10.3847/1538-4357/aafad1}, \href
  {https://ui.adsabs.harvard.edu/abs/2019ApJ...872...13W} {872, 13}

\bibitem[\protect\citeauthoryear{{Y{\`e}che}, {Palanque-Delabrouille}, {Baur}
  \& {du Mas des Bourboux}}{{Y{\`e}che} et~al.}{2017}]{Yeche17Neutrino}
{Y{\`e}che} C.,  {Palanque-Delabrouille} N.,  {Baur} J.,   {du Mas des
  Bourboux} H.,  2017, \mn@doi [\jcap] {10.1088/1475-7516/2017/06/047}, \href
  {http://adsabs.harvard.edu/abs/2017JCAP...06..047Y} {6, 047}

\bibitem[\protect\citeauthoryear{{Zaldarriaga}, {Hui}  \&
  {Tegmark}}{{Zaldarriaga} et~al.}{2001}]{Zaldarriaga01}
{Zaldarriaga} M.,  {Hui} L.,   {Tegmark} M.,  2001, \mn@doi [\apj]
  {10.1086/321652}, \href {http://adsabs.harvard.edu/abs/2001ApJ...557..519Z}
  {557, 519}

\makeatother
\end{thebibliography}

\appendix
\section{Power spectrum without masking metal absorption lines}\label{A}
\begin{figure*}
\includegraphics[width=0.995\textwidth,height=\textheight,keepaspectratio]{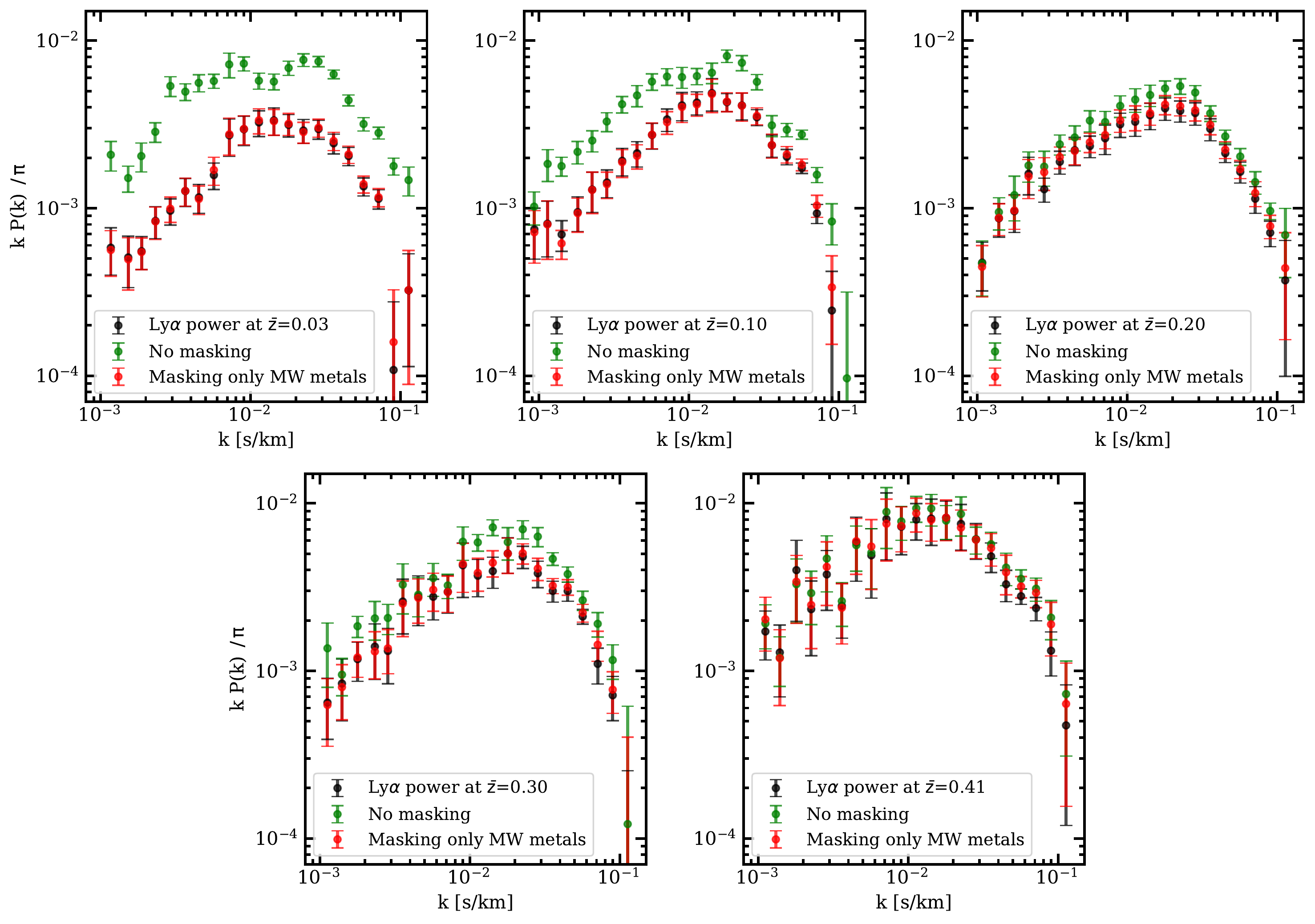}
\caption{  Illustrating the effect of masking different metal lines on the power spectrum. Black 
points show our power spectrum measurements as also shown in Fig.~\ref{fig3}. Power without 
masking any metals (green points) is systematically higher at all scales. At lower-$z$, difference 
arises mainly because of Milky-way metals. This is illustrated by power spectrum shown in red 
points which is obtained by masking only the Milky-way metals. A small difference between red and 
black points shows the effect of masking intervening metals. The intervening metals have 
insignificant impact on the Lyman-$\alpha$ power spectrum at low-$z$. On the other hand, Milky-way 
metal lines are easy to identify which makes any incompleteness in intervening metal 
identification unimportant for calculating the power spectrum.
} 
\label{figA1}
\end{figure*}

While preparing data for the power spectrum calculation, we have properly 
masked the metal lines originating  from Milky-way and intervening metals 
as well as other contaminations (see Section 2). This is an important
procedure to correctly calculate the power spectrum. 
However, to estimate the relevance of different metal contamination, in Fig.~\ref{figA1} we show the 
power spectrum calculated 
without masking any metal absorption lines (green points) and masking only Milky-way metal lines (red
points) in the Lyman-$\alpha$ forest. However, note that we mask the spectral gaps and all other 
emission lines, which appear mainly in our lowest redshift bin ($\bar z = 0.03$). As expected, when 
metal lines are not masked the power is systematically higher 
because of the extra fluctuations in the flux introduced by metal lines. However the difference in 
power is larger at lower redshift mostly 
because of strong metal contamination arising 
from the ISM of the Milky-way and many of the absorption lines are correlated as they originate from 
the same source and with same wavelength separation.  The intervening metals have insignificant 
impact on the Lyman-$\alpha$ power spectrum at low-$z$ (the difference between red and black points 
in Fig~\ref{figA1}). On the other hand, Milky-way metal lines are easy to identify which makes any 
incompleteness in intervening metal identification unimportant for calculating the power spectrum.

\section{Effect of masking on the power-spectrum}\label{B}
\begin{figure*}
\includegraphics[width=0.995\textwidth,height=\textheight,keepaspectratio]{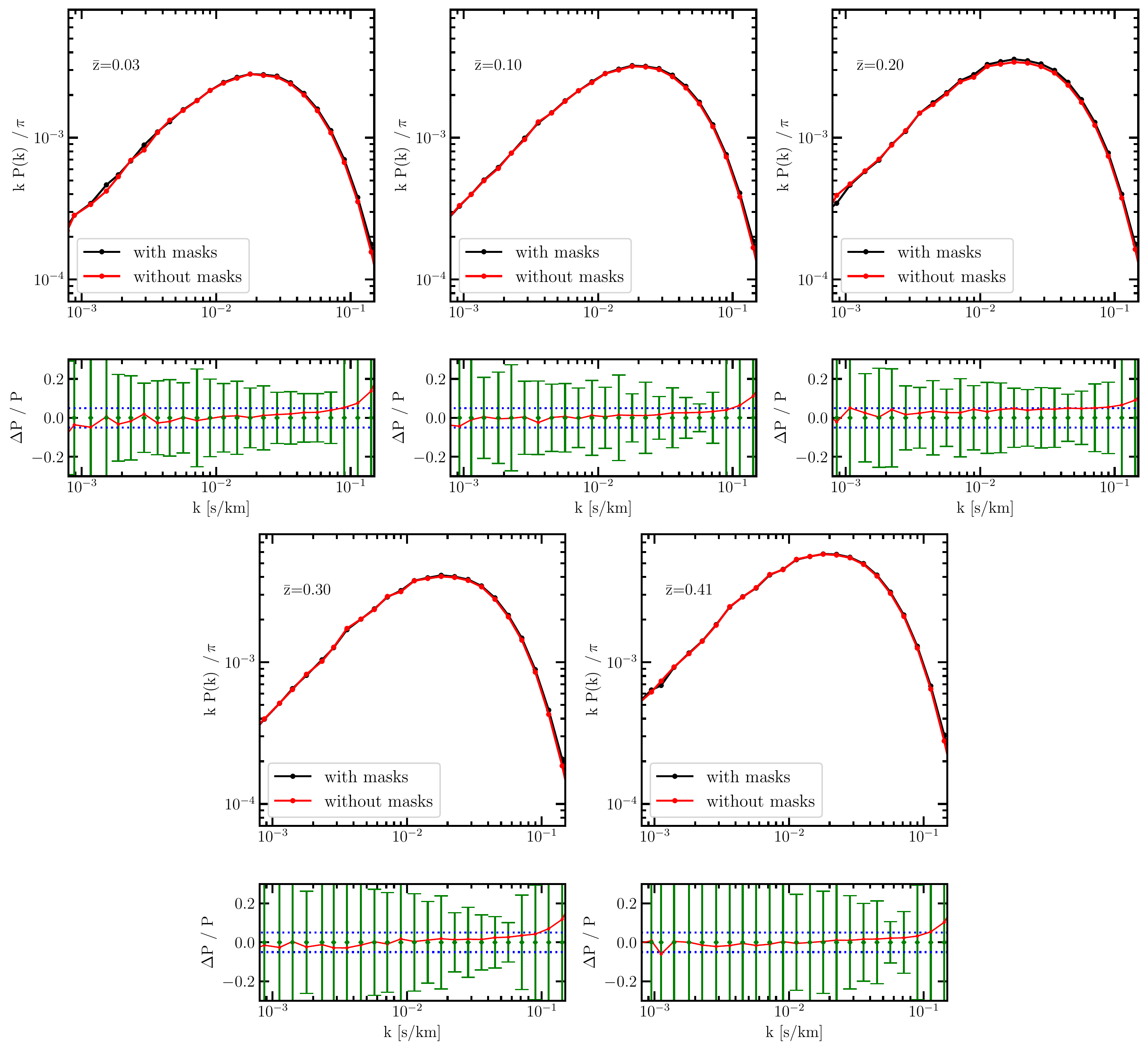}
\caption{ Effect of masking on the power spectrum. We use mock spectra generated from our forward 
models (infinite signal and resolution) and calculate the power spectrum for two cases. First by 
applying exactly same masks as in data to the forward model spectra (black-curves) and second where 
no masks are applied (red curves). The both curves match with each other with $<5\%$ difference as 
shown in the bottom panels where 5\% change is marked with blue dotted line. The green error-bars 
show out fractional errors from the measurements. The difference is small enough to ignore and opt 
out of the masking correction.}
\label{figA2}
\end{figure*}

To check if the spectral masking is introducing any significant contamination
in the power-spectrum we calculate the power spectrum from our forward models (using
$\Gamma_{\rm H \,I}$ values consistent with our measurements) but without using 
any masking on the Lyman-$\alpha$ forest. In these forward models for the 
purpose of only studying the effect of masking, we are using infinite resolution 
noiseless spectra. In Fig.~\ref{figA2} we compare the resulting power spectrum 
with and without masking the Lyman-$\alpha$ forest. The bottom panel shows a 
fractional difference between these. The maximum change in relevant $k<0.1$ s 
km$^{-1}$ values is no more than 5\% at any redshift. For comparing with the 
actual uncertainties in the power spectrum, we also plot the fractional errors 
in the bottom panel. The figure clearly shows that the fractional change in the 
power-spectrum because of masking is less than the errors on the actual power 
spectrum measurements. This motivates us not to perform the masking 
corrections on the measurements as done in \citet{Walther18a}.

\section{Effect of the COS LSF on the power-spectrum}\label{C}
\begin{figure*}
\includegraphics[width=0.995\textwidth,height=\textheight,keepaspectratio]{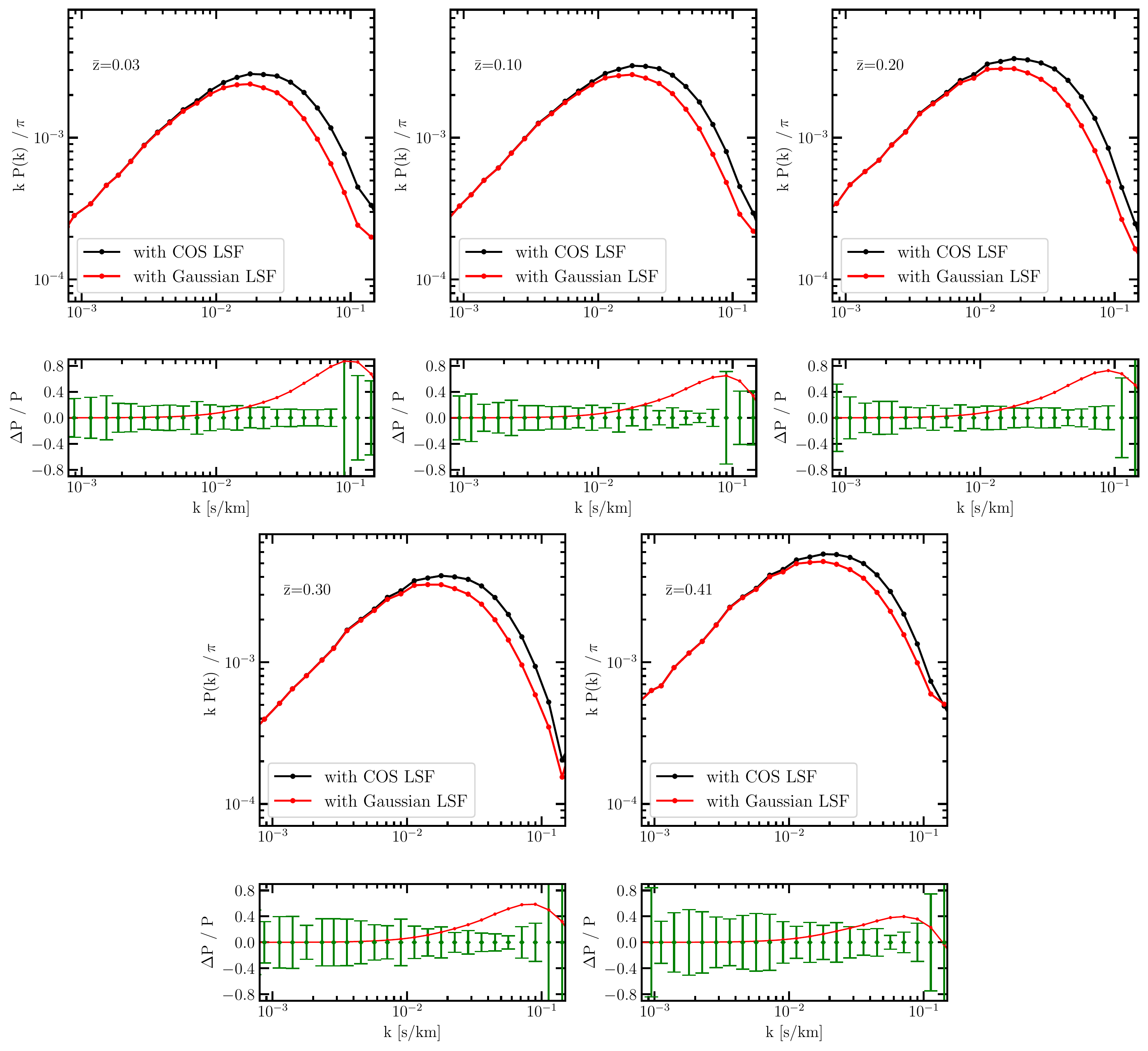}
\caption{Effect on the power spectrum of incorrectly using the Gaussian LSF instead of COS LSF that 
has broad non-Gaussian wings. We use our forward models generated using the COS LSF and calculate the
power spectrum with a Gaussian approximated to the COS LSF (red curves). This power spectrum is 
compared with that calculated using correct COS LSF (black curves). The bottom panels show the 
percentage difference between these compared with the fractional errors estimated from our 
measurements. The differences are too large to ignore. Therefore, it is never recommended to 
approximate the COS LSF with a Gaussian, at least for power spectrum calculations.}  
\label{figA3}
\end{figure*}

The COS LSF is quite different from a Gaussian as it exhibits strong non-Gaussian wings. In the power
spectrum calculation 
the LSF appears in the window function correction (see Eq. 1) and it is 
also important to generate the forward models. To see how much the power 
spectrum can be affected if one uses a Gaussian LSF with a similar resolution to 
COS we perform the following analysis. We use mock spectra generated from
our forward models (at 
$\Gamma_{\rm H \,I}$ values consistent with our measurements) generated
using the correct COS LSF but while calculating the power spectrum we use a Gaussian 
LSF. The resulting power spectrum (red curves) is shown in the Fig~\ref{figA3} 
along with that obtained by using the correct COS LSF (black curves). 
The amplitude of the power spectrum in the former is smaller at large $k$ 
values and the differences, as shown in the bottom panels, can be as large as 
80\%. This is much too large as compared to the actual uncertainties from our 
measurements. Although the differences decrease at higher redshifts, because of 
small sample sizes, the discrepancy is too significant to ignore. This exercise 
shows that it is never a sound choice to approximate the COS LSF as a Gaussian. 

\bsp	
\label{lastpage}
\end{document}